\documentclass[10pt,aps,prl,twocolumn,superscriptaddress,floatfix]{revtex4-1} %

\usepackage{graphicx}
\usepackage{graphics}
\usepackage{amsmath}
\usepackage{amsfonts, amssymb}
\usepackage{float}
\usepackage{subfig}
\usepackage{tabularx}
\usepackage[normalem]{ulem}
\usepackage{color}
\usepackage[usenames,dvipsnames]{xcolor}
\usepackage[version=3]{mhchem}
\usepackage{multirow}
\usepackage{lipsum}
\usepackage[T1]{fontenc} 
\definecolor{gCCOC}{RGB}{0,194,0}
\definecolor{oCH2}{RGB}{255,128,0}
\definecolor{bCH3}{RGB}{0,0,255}
\definecolor{rCO}{RGB}{255,94,255}
\definecolor{gNH2}{RGB}{0,255,0}
\definecolor{vCO}{RGB}{194,64,255}
\definecolor{rNH2}{RGB}{220,32,50}
\definecolor{rsp3}{RGB}{194,0,0}
\definecolor{yCO}{RGB}{161,161,0}

\begin{document}
\title{Spectroscopic studies of the mechanism of reversible photodegradation
of 1-substituted aminoanthraquinone-doped polymers}
\author{Sheng-Ting Hung}
\email{sth@wsu.edu}
\affiliation{Department of Physics and Astronomy, Washington State University,
Pullman, WA 99164-2814}
\affiliation{Department of Chemistry, University of Leuven, Leuven, Belgium B-3001}
\author{Ankita Bhuyan}
\affiliation{Department of Physics and Astronomy, Washington State University,
Pullman, WA 99164-2814}
\author{Kyle Schademan}
\affiliation{Department of Physics and Astronomy, Washington State University,
Pullman, WA 99164-2814}
\author{Joost Steverlynck}
\affiliation{Department of Chemistry, University of Leuven, Leuven, Belgium B-3001}
\author{Matthew D. McCluskey}
\affiliation{Department of Physics and Astronomy, Washington State University,
Pullman, WA 99164-2814}
\author{Guy Koeckelberghs}
\affiliation{Department of Chemistry, University of Leuven, Leuven, Belgium B-3001}
\author{Koen Clays}
\affiliation{Department of Physics and Astronomy, Washington State University,
Pullman, WA 99164-2814}
\affiliation{Department of Chemistry, University of Leuven, Leuven, Belgium B-3001}
\author{Mark G. Kuzyk}
\affiliation{Department of Physics and Astronomy, Washington State University,
Pullman, WA 99164-2814}
\date{\today}

\begin{abstract}
The mechanism of reversible photodegradation of 1-substituted aminoanthraquinones
doped into poly(methyl methacrylate) and polystyrene is investigated. 
Time-dependent density functional theory is employed to
predict the transition energies and corresponding oscillator strengths of the proposed reversibly- and irreversibly-damaged dye species. Ultraviolet\textendash visible
and Fourier transform infrared (FTIR) spectroscopy are used to characterize which species are present. FTIR spectroscopy
indicates that both dye and polymer undergo reversible photodegradation
when irradiated with a visible laser. These findings suggest that
photodegradation of 1-substituted aminoanthraquinones doped in polymers 
originates from interactions between dyes and photoinduced thermally-degraded polymers, and the metastable product may recover or further degrade irreversibly.

\vspace{1em}

\end{abstract}

\maketitle

\vspace{1em}

\section{Introduction}
Organic materials have a broad range of applications 
such as high-resolution fluorescence microscopy \cite{White1987,Denk1990,Zipfel2003,Huang2009,Rego2012,Linde2012,Horton2013},
second harmonic generation (SHG) microscopy \cite{Reeve2009,Reeve2010,Meulenaere2012,Lopez2013,Redon2014},
dye sensitized and polymer solar cells \cite{Hoppe2004,Guness2007,Hagfeldt2010,Li2012,Congreve2013},
solid state dye/organic lasers \cite{Costela2003,Chenais2012,Chen2014}, and
organic light emitting diodes \cite{Kulkarni2004,Kamtekar2010}, to name a few. Photostability
of organic compounds and polymers is often a requirement for applications
incorporating light-matter interaction \cite{Denk1990,Zipfel2003,Santra2004,Donnert2006,Reeve2009,Hoebe2007,Chen2014,Huang2009,Linde2012,Agrell2003,Ahmad2002,Costela2003,Chenais2012,Jorgensen2012,Tournebize2013,Cao2014}.
When a material undergoes photodegradation, its characteristic properties
deteriorate over time, which is referred to as decay; the reverse change
in the characteristic properties of the material is referred to as
recovery. Though photodegradation is often irreversible, the recovery
process has been observed from a large variety of materials, typically
involving polymers and often together with dyes, with various experimental
techniques when the photodegraded materials are kept in dark for
a long enough time, typically hours to days \cite{Peng1998,Przhonska1998,Howell2002,Singh2002,Kobrin2004,Biancardo2006,Katz2006,Zhu2007,DesAutels2009,
Seemann2009,Anderson2011a,Anderson2013b,Ayesta2014,Cao2014,Anderson2015}.

Amplified spontaneous emission (ASE) of disperse orange 11 (DO11)
doped in poly(methyl methacrylate) (PMMA) bulk sample was observed
to fully recover in the dark about 40 hours after photodegradation
when irradiated with a 532 nm second harmonic Nd:YAG picosecond laser
\cite{Howell2002}. This self-healing phenomenon was also observed
in various anthraquinone derivatives doped in PMMA and polystyrene
(PS) thin films \cite{Anderson2011a,Anderson2013b} and DO11 doped
in MMA-styrene copolymers thin films \cite{Anderson2013b,Hung2012}
probed with transmittance image microscopy and ASE. The photodegraded
thin film samples are often observed to recover partially, which suggests
there exists both reversible and irreversible photodegradation.
The lack of evidence for linear dichroism during photodegradation measurements eliminates orientational hole burning
as the mechanism causing reversible photodegradation \cite{Embaye2008}.
Spatially resolved ASE and fluorescence \cite{Ramini2011}, spatial
imaging \cite{Anderson2011b} and ultraviolet\textendash visible (UV-Vis)
spectroscopy \cite{Hung2015} studies indicate that diffusion/back
diffusion is not responsible for reversible photodegradation. While
some phenomenological kinetic models have been proposed based on quantitative
studies of reversible photodegradation in dye-doped polymer (dye/polymer) samples 
including AF455/PMMA and DO11/PMMA, a decade of research using various experimental techniques  \cite{Zhu2007,Embaye2008,Ramini2012a,Ramini2012b,Ramini2013,Anderson2013a,Anderson2013b,Anderson2014a},
they have failed to elucidate the underlying mechanism.

Several hypotheses of the mechanism responsible for reversible photodegradation
in DO11/PMMA have been proposed, including intramolecular proton transfer and dimer formation
\cite{Embaye2008}, twisted intramolecular charge transfer \cite{Westfall2012},
and domain-assisted reversible photodegradation \cite{Ramini2012b,Kuzyk2012a,Kuzyk2012b},
none have been experimentally proven.
These hypotheses, together with anion formation \cite{Hocking1982,Linde2011}
and protonation \cite{Gordon1983,Zarzeczanska2014} of dye molecules
observed in similar systems, have been investigated and the results
suggest that they are unlikely responsible for reversible photodegradation
of 1-substituted aminoanthraquinone derivatives doped in PMMA and
PS \cite{Hung2015}. 

This study aims to understand the underlying mechanism of reversible
and irreversible photodegradation of 1-substituted aminoanthraquinone-doped
polymers by combining Gel permeation chromatography (GPC), time-dependent density 
functional theory (TD-DFT) calculations, UV-Vis spectroscopy and FTIR spectroscopy to characterize chemical species and changes in molecular structures during photodegradation and recovery. 
Most research reported in the literature on 1-substituted aminoanthraquinone 
derivatives focus on 1-aminoanthraquinone (1AAQ) instead of DO11, the molecule
studied most extensively in our laboratory. The fact that 
reversible photodegradation has been observed in both 1AAQ and DO11
doped in PMMA matrices suggests that photoinduced reactions between
1AAQ or DO11 and the polymer host are the same process \cite{Hung2015}. Therefore, both 1AAQ and DO11 are utilized
in this study and the results obtained from each molecule are assumed
to be applicable to each other.


\section{Background}
Anthraquinone exhibits a weak optical absorption band at 405 nm \cite{Bruce1974,Gordon1983},
with an extinction coefficient about 60 cm\textsuperscript{-1}M\textsuperscript{-1}
(in methanol) which is often undetected \cite{Gordon1983}. Aminoanthraquinones,
however, possess moderate to strong absorption bands in the visible regime,
which are absent in anthraquinone, and have been assigned to intramolecular
charge transfer (ICT) between the amine group and the carbonyl groups
\cite{Inoue1972,Inoue1973}. For example, DO11 dissolved in MMA is
observed to show an absorption peak at 471 nm with an extinction
coefficient of $8.043\times10^{3}$ cm\textsuperscript{-1}M\textsuperscript{-1}. The calculated electron density of
1AAQ is also observed to increase in the unsubstituted ring and in the
carbonyl groups in the first excited state relative to the ground
state \cite{Gordon1983}. This indicates that the carbonyl groups
are not the only electron acceptors \cite{Gordon1983}. This observation is consistent
with the results reported by Inoue et al. that the calculated electron
density of 1AAQ of the first excited state is increased in the carbonyl
groups, the center ring and the unsubstituted ring, though they only
mentioned the carbonyl groups \cite{Inoue1972}.

Photocycloaddition of 1AAQ to olefines, including styrene, was observed
with visible light irradiation provided by an optically filtered ($\lambda>420$
nm) 300 W high-pressure mercury lamp at 0 $^{\circ}$C \cite{Inoue1979,Inoue1982b}.
In this photochemical reaction, the carbonyl group adjacent to the
amine group of 1AAQ reacts with a diene (or olefine) under exposure
of visible light to form a corresponding oxetane. The structures of
some reaction products were confirmed by IR, NMR, and mass spectrometry
\cite{Inoue1979,Inoue1982b}.

Although there was no photocycloaddition observed between 1AAQ and
monoenes \cite{Inoue1979,Inoue1982b}, the reason was not mentioned. A possible explanation is that the light source
used in the experiment did not provide enough intensity to show a
measurable reaction rate, but if the solution is irradiated with a
higher-powered light source such as a laser, the reaction rate might
be increased above the detection threshold. Another hypothesis is
that the visible light source did not provide enough energy to overcome
the activation energy for photocycloaddition of 1AAQ with monoenes
at 0 $^{\circ}{\rm C}$. However, when irradiating the solution with
a more intense light source such as a laser, the energy density in
the vicinity of the irradiated dye molecules might be temporarily
high due to nonradiative relaxation of photo-excited dye molecules, thus,
overcoming the activation energy of photocycloaddition. Thus, photocycloaddition
may occur between 1-substituted aminoanthraquinones and olefines in
general, including monoenes such as methyl methacrylate (MMA) when
exposed to a laser.

Photocycloaddition was proposed to involve excited complex (exciplex)
formation between an excited ICT state of 1AAQ and an olefine \cite{Inoue1982c}.
The photocycloaddition was unaffected by oxygen, and the reaction
products were unstable in an environment of carbon dioxide, acid,
light, and heat \cite{Inoue1979,Inoue1982b}. Most importantly, the
reaction products were found to gradually decompose into other compounds
including 1AAQ itself at 30 $^{\circ}{\rm C}$ \cite{Inoue1979}.
The return of 1AAQ from the photocycloaddition products could be the
recovery process observed in 1AAQ-doped polymers if photocycloaddition
occurs between 1AAQ and the polymer hosts or fragments of thermally
degraded polymers, including depolymerized monomers due to locally
accumulated heat via nonradiative relaxation of excited dye molecules.

Thermal degradation of polymers has been studied extensively for several
decades, and the dominant products of thermal degradation of PS and
PMMA are known to be their monomers, styrene and MMA, respectively
\cite{Grassie1985,McNeill1990,Manring1991}. Mechanisms of thermal
degradation such as depolymerization, scission of side chains, and
dissociation of the polymer backbones take place depending on the
environment, temperature, molecular weight, chain end groups, chain
configuration, polymerization condition etc. \cite{Grassie1985,McNeill1990,Peterson2001,Ferriol2003,Stoliarov2003,Manring1988,Manring1989a,Manring1989b,Manring1991,Kashiwagi1985,Kashiwagi1986,Hirata1985,Holland2001,Holland2002,Hu2003}
As such, the activation energy of thermal degradation varies widely.
It is generally found to be between 1.87 eV and 3.34 eV for PS in
an inert atmosphere or vacuum \cite{Peterson2001}, and from 1.23
eV to 3.55 eV for PMMA in an inert atmosphere \cite{Hirata1985,Kashiwagi1985,Manring1988,Manring1991,Holland2001,Holland2002,Ferriol2003,Stoliarov2003}.
The photon energies of lasers used in this study are 2.33 eV and 2.54
eV, which are within the reported range of activation energies for
thermal degradation of PS and PMMA.

The fluorescence quantum yield of 1AAQ in several organic solvents
was reported to be less than 10\% \cite{Inoue1982a,Dahiya2006}. Thus,
most of the photon energy absorbed by 1AAQ molecules is lost via nonradiative
relaxation, leading to an accumulation of heat centered on the 1AAQ molecules.This local buildup of heat can cause thermal
degradation of nearby polymer chains.

Photoinduced damage in dye-doped polymer matrices has been studied
in various systems for different applications \cite{Rabek1974,Mita1984,Mita1988,Kaczmarek2000,Mita1990,Chiantore2000,Fukumura1993,Egami1998,Silverstein1998a,Popov1998}.
Sensitized photodegradation and photooxidation of polymers can be
induced by photoinitiators, in which polymer degradation or oxidation
can be initiated by free radicals originating from photodegradation
of sensitizers; or by photosensitizers, in which excited sensitizers
transfer energy to the polymer or oxygen to initiate photodegradation
or photooxidation \cite{Rabek1974}. Studies on photooxidation or
photodegradation of various sensitized polymers have been reported
including sensitized PS \cite{Rabek1974,Mita1984,Mita1988,Kaczmarek2000}
and sensitized PMMA \cite{Mita1990,Chiantore2000}. Laser ablation
of polymers can be induced with the assistance of photosensitizers
using visible lasers at wavelengths of 351 nm \cite{Fukumura1993},
488 nm \cite{Egami1998}, and 532 nm \cite{Silverstein1998a}. Fukumura
et al. proposed that the photon energy absorbed by anthracene, the
sensitizer, is converted into thermal energy in PS, the polymer host,
causing thermal decomposition \cite{Fukumura1993}. In a study of
photodestruction of a solid-state dye laser composed of Rh6G-chloride
dye doped in a modified PMMA gain medium, Popov attributed photobleaching
of excited-state dye molecules to the permanent degradation of the
lasing efficiency, and assigned the formation of carbon (char) in
the polymer matrix to the polymer decomposition due to accumulated
heat transferred from photoexcited dye molecules via direct vibrations
and nonradiative relaxation of their singlet levels \cite{Popov1998}.

\subsection{Hypothesis}
Given the above reasoning that is guided by observations, we propose
that a series of photothermally-induced chemical reactions (PTCR) between
dye and polymer is a potential mechanism responsible for reversible
and irreversible photodegradation of 1AAQ (and DO11) doped in PMMA
and PS as follows:

1AAQ (and DO11) undergoes ICT when excited with visible light, and
the absorbed photon energy can be transferred to nearby polymer chains
via nonradiative relaxation of the excited ICT singlet states. The
transferred energy locally heats the polymer around the excited dye
molecules, which leads to thermal degradation of the polymer including
depolymerization, scission of side chains, and dissociation of polymer
backbones. Since monomers are the major products of thermally degraded
polymers (PS and PMMA), an excited dye molecule can undergo photocycloaddition
with a monomer in its vicinity to form the reversibly photodegraded
dye species. In addition to monomers, there are other decay products
such as $\cdot$\ce{CH3}, $\cdot$\ce{COOCH3}, $\cdot$phenyl, 
and polymer chain radicals etc., which can attack ground- and excited-state
dye molecules and reversibly-damaged dye species causing irreversible
damage to dye molecules. While reversibly-damaged dye species may
gradually recover back to the original dye molecule with a monomer
left behind in the polymer matrix, it may also further decompose to
an irreversibly-damaged dye species or dye radical or other fragments
of radicals. Thus, the ``recovered'' and thermally degraded monomers
and radical fragments can react with each other or with decomposed
polymer fragments, polymer chains and unsaturated polymer chain ends
causing recovery of the polymer or changes in the polymer including
the formation of small molecular weight polymer chains, cross linked
polymer chains, and perhaps new species of polymer segments and chains.

\begin{figure}[h]
\begin{centering}

\includegraphics[scale=0.62]{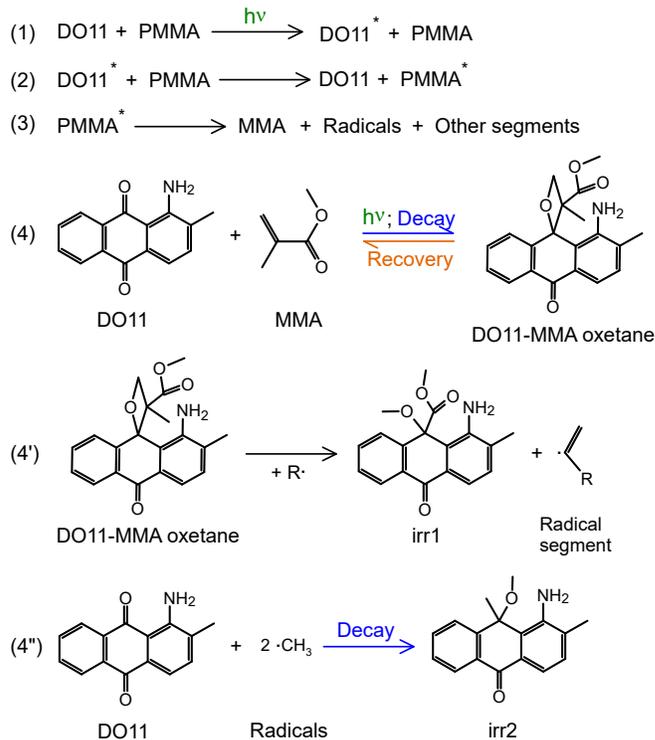}
\par\end{centering}

\centering\caption{Reversible and irreversible photodegradation of DO11/PMMA. The asterisk indicates photoexcited DO11 or heated PMMA through nonradiative energy transfer from excited DO11. Reversible and irreversible degradation in DO11 starts after thermal degradation of PMMA (step (3)).Non-radical segments can also undergo thermal degradation similar to step (2) and (3) resulting in radicals. The reversibly-damaged species may undergo recovery (back reaction of step (4)) or decompose into irreversibly-damaged species and radical segments (step(4')). ``irr1'' and ``irr2'' are possible irreversibly-damaged dye species illustrated in Figure\ref{fig:PChemSpecies}. There can be other irreversibly-damaged dye species depending on DO11's nearby radicals and polymer segments.}
\label{fig:PChemReaction}

\end{figure}

The PTCR hypothesis is illustrated with DO11/PMMA in Figure \ref{fig:PChemReaction}.
In this mechanism, dye undergoes (photo)degradation after the thermal
degradation of polymer chains. The reversible degradation of dye requires
at least two photons initially, but there can be multiple monomers
available in the vicinity of a dye molecule once the degradation of
the polymer takes place, so the reversible degradation does
not necessarily require two photons. On the other hand,
the irreversible degradation of dye may be caused by radicals produced
from at least one-photon-induced thermal degradation of the polymer.

\begin{figure}[h]
\begin{centering}
\includegraphics[scale=0.4]{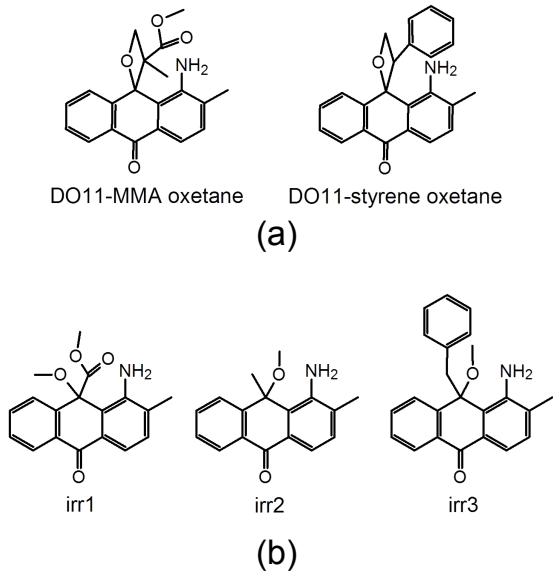}
\par\end{centering}

\centering\caption{\label{fig:PChemSpecies}(a) Proposed reversibly-damaged DO11 biproducts
made by photocycloaddition with MMA (left) and styrene (right). (b)
Illustration of possible irreversibly-damaged DO11 species.}
\end{figure}

Reversibly-damaged DO11 in PMMA and PS is proposed to be oxetanes as shown in Figure \ref{fig:PChemSpecies}(a).
Irreversibly-damaged DO11 in PMMA and PS are illustrated in
Figure \ref{fig:PChemSpecies}(b) but other species may form due to 
the various radicals formed upon thermal degradation of
polymer chains. The important point is that the carbonyl group adjacent
to the amine group is damaged, and the carbon and oxygen atoms singly
bond to thermally-degraded polymer fragments or polymer chains as
will be rationalized later in the discussion.

\section{Computational method}
Ultraviolet-visible (UV-Vis) spectra of various organic molecules
including anthraquinone derivatives have been calculated using TD-DFT
and the results are found to be in good agreement with
the observed spectra if the appropriate basis set of wavefunctions and functionals
are chosen \cite{Jacquemin2004,Jacquemin2007,Fabian2010,Jacquemin2011,Jacquemin2012a}.
In order to obtain a qualitative estimate of the absorption spectra of
possible photodegraded dye molecules within a reasonable computing
time, a method that produces reasonably good predictions
for visible spectra of anthraquinone derivatives is adopted \cite{Jacquemin2004}
in this study. 

Gaussian 03 was used to perform the TD-DFT calculations.
Geometry optimization of pristine DO11 and all possible damaged DO11
species in Figure \ref{fig:PChemSpecies} were carried out with the
6-31G(d,p) basis set and the B3LYP functional for each species in
vacuum. TD-DFT calculations were carried out with the optimized ground
state geometry using the same basis set and functional, and the polarizable
continuum model (PCM) was used to account for solvent effects. Photodegradation
experiments were performed in PS and PMMA polymer matrices that have
dielectric constants of 2.6 and 2.6-3.1 \cite{CRC1978}, respectively,
which are somewhat close to toluene's value of 2.38 \cite{CRC1978}. Thus,
toluene was selected as the PCM solvent in TD-DFT calculations.


\section{Experimental method}

\subsection{Sample preparation}
1AAQ and 1-amino-2-methylanthraquinone (DO11) with a purity of 97\%
and 95\%, respectively; and methyl methacrylate (MMA) and styrene
were purchased from Aldrich. MMA and Styrene were purified using two
column flasks (one for each) filled with alumina powder to remove
the inhibitors that prevent monomers from polymerization. Commercially
available PMMA (MW $=120,000$) purchased from Aldrich was also used
in this study.

\paragraph{Thin-film samples prepared from monomers}
To make dye-doped PMMA samples, dye and purified MMA monomer were
mixed in proportions to obtain the desired dye concentration and were
sonicated for 30 minutes. After sonication, both butanethiol (chain
transfer agent) and tert-butyl peroxide (initiator) were
added to the solutions in the proportion of 3.3 $\mu$l per ml MMA,
and were sonicated for another 30 minutes. The sonicated solutions
were filtered with 0.2 $\mu$m syringe filters and placed in an oven
at 95 $^{\circ}{\rm C}$ to initiate the polymerization reaction for
at least 2 days resulting in dye-doped polymers. An appropriate volume
of polymerized sample was pressed between two 2.5 cm $\times$ 2.5
cm glass substrates at 140 $^{\circ}{\rm C}$ (well above the glass
transition temperature where the polymer flows) with an uniaxial pressure
of $110\pm10$ psi for 90 minutes to make a thin film. The pressure
was gradually reduced while the sample was cooling at an average
rate about 1.5 $^{\circ}{\rm C}$/min.

For dye-doped PS samples, the same procedure as described above was
used, but with 4.2 $\mu$l of initiator added to 1 ml of styrene. The filtered
solutions were placed in an oven at 95 $^{\circ}{\rm C}$ for 4 days
for complete polymerization. The glass transition temperature of dye-doped
PS samples is lower than dye-doped PMMA samples, so the temperature
for pressing thin films was reduced to 120 $^{\circ}{\rm C}$.

Thin-film samples obtained by this method have thicknesses in the
range of 20 - 120 $\mu$m depending on the amount of polymerized sample
pressed between the glass slides.

\paragraph{Thin-film samples prepared from PMMA}
Appropriate amounts of dye and PMMA in the desired ratio were dissolved
into a solution composed of 33\% $\gamma$-butyrolactone and 67\%
propylene glycol methyl ether acetate (PGMEA) with 10\% solute and
90\% solvent by weight. The solution was stirred for 2 to 3 days
to dissolve the dye and polymer, then filtered with a 0.2 $\mu$m
syringe filter. The filtered solution was then spin-coated on 2.5
cm $\times$ 2.5 cm glass substrates at 600 rpm for 90 seconds to 
make films for linear absorption spectroscopy measurements. The filtered
solution was also spin-coated on a 2.5 cm $\times$ 2.5 cm silicon
wafer at 3000 rpm for 50 seconds to make films for performing FTIR experiments.
All spin-coated samples were stored in a vacuum oven at 100 $^{\circ}{\rm C}$
for 90 minutes; then, the heater was turned off and the samples were
kept in vacuum over night. 

The thickness of the films made by spin-coating 
on a glass slide is about 0.6 $\mu$m estimated by the absorbance
of the 1AAQ/PMMA thin film \cite{Hung2015}.
The thickness of films made by spin-coating on a silicon substrate is not
measured, but based on the fact that the spin speed
was 5 times faster than the sample spin-coated on a glass slide the film's thickness is expected to be less than 0.6 $\mu$m.

\subsection{GPC}
PMMA purchased from Aldrich and DO11/PMMA polymerized from DO11 dissolved in MMA 
monomers with concentration 9 g/L were used in the GPC experiment. MMA and DO11 were 
also tested with GPC as a control. MMA, DO11, PMMA and DO11/PMMA samples were 
dissolved in Tetrahydrofuran (THF), filtered through a 20 $\mu$m filter and injected. A 
Shimadzu 10A apparatus with a PLgel 5 $\mu$m mixed-D type column and a refractive index 
detector (RID) and UV-Vis spectrometer were used as detectors.

\subsection{UV-Vis spectroscopy}
\begin{figure}[h]
\begin{centering}
\includegraphics[scale=0.47]{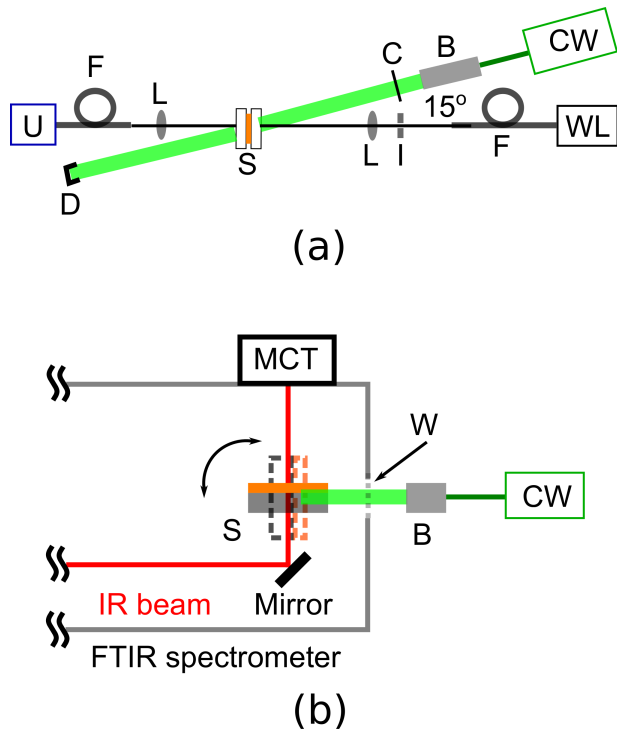}
\par\end{centering}

\protect\caption{\label{fig:AbsFTIRSetup}(a) UV-Vis spectroscopy setup.
(b) FTIR spectroscopy setup. The pump beam is a cw laser and is expanded
5 times in diameter. (CW) CW Laser. (B) 5$\times$ beam expander.
(C) Shutter. (D) Beam dump. (F) Optical fiber. (I) Iris. (L) Convex
lens. (S) Sample. (U) Spectrometer. (WL) White light source. (W) Quartz
window. (MCT) the MCT detector.}
\end{figure}

The setup for the UV-Vis spectroscopy experiments is schematically shown
in Figure \ref{fig:AbsFTIRSetup}(a). A continuous wave (cw) argon
ion laser providing the pump beam at a wavelength of 514 nm and a cw solid
state laser providing the other pump beam at a wavelength of 532 nm were used
to induce photodegradation. Two light sources
were used for absorption measurements: an Ocean Optics PX2 Xenon pulsed
lamp and an Ocean Optics LS1 tungsten halogen lamp. The spectrometer
was an Ocean Optics Model SD2000. The angle between the cw pump beam and
the white light path was about 15$^{\circ}$. The 532 nm pump laser
beam passed a $5\times$ beam expander to make a 1/e diameter of
7.5 mm. The argon ion laser beam was not expanded and the 1/e diameter
of 514 nm was 1.01 mm. The white light was focused to a diameter of 0.6
mm. The white light probe was centered on the cw pump beam at the sample.
The sample was mounted on a translation stage, which allowed 
the reference spectrum to be retaken through air in proper time intervals during
recovery to ensure that the change of absorbance is not due to drift of
the white light intensity.

While the cw laser was causing photodegradation in the sample, the
absorption spectra were recorded in proper time intervals with the
pump laser temporarily blocked. After the desired irradiation time
for photodegradation, the cw laser was turned off and absorption spectra
were recorded in preset time intervals to monitor recovery.

The absorption spectra of irreversibly-damaged dye species can be
approximated by irradiating the sample for a long enough time so that
the absorption spectrum no longer changes. Thus, a DO11/PS sample
of concentration 9 g/L was irradiated with the 532 nm wavelength laser
at a peak intensity of 35.01 W/cm\textsuperscript{2} for 90 minutes, and
a DO11/PMMA sample of concentration 9 g/L prepared from MMA monomers
was irradiated with the 514 nm wavelength laser at a peak intensity of 34.91
W/cm\textsuperscript{2} for 180 minutes to estimate the absorption
spectra of irreversibly-damaged dye species.

\subsection{FTIR spectroscopy}
IR spectra were obtained from a DA8 Bomen FTIR spectrometer equipped
with a water-cooled globar broadband light source (200 to 10000 cm\textsuperscript{-1}),
a KBr beam splitter (450 to 5000 cm\textsuperscript{-1}), and a liquid
nitrogen cooled MCT detector (400 to 5000 cm\textsuperscript{-1}).
The pressure inside the FTIR spectrometer was kept below 0.2 torr throughout
the entire experiment. The aperture was set at 10.0 mm, the speed of
the moving mirror was 0.5 cm/s, and the resolution was 4 cm\textsuperscript{-1}.
The customized sample holder has 3 holes allowing light to pass
and each of them is 5 mm in diameter, so three spin-coated
(dye-doped) polymer samples can be loaded simultaneously. The 532 nm pump
beam generated from a cw solid state laser was expanded with a $5\times$
beam expander and sent into the sample chamber through a quartz window
resulting in a 7.5 mm diameter pump beam at the sample.
The peak intensity of the expanded pump beam was approximately 2.09
W/cm\textsuperscript{2}. The setup for the FTIR spectroscopy experiments
is schematically shown in Figure \ref{fig:AbsFTIRSetup}(b). The dye-doped
polymer sample can be rotated 90$^{\circ}$ for laser irradiation,
and rotated back to the original position for taking IR spectra after
irradiation. The quartz window was covered and the laser beam was blocked
while the IR spectra were taken.

For FTIR experiments, a plain silicon wafer and a PMMA film on
a silicon wafer were prepared as controls. A 1AAQ/PMMA film
on a silicon wafer was prepared for probing chemical
bonds in dye and the host polymer during reversible photodegradation.
To minimize the uncertainty, all three samples were fixed on the sample
holder and placed in the vacuum chamber for the entire experiment.

All IR spectra were acquired at room temperature. IR spectra taken
before irradiation were averaged over 6000 scans, which took about
66 minutes. The 1AAQ/PMMA sample was irradiated for 30 minutes. IR
spectra of 1AAQ/PMMA acquired after irradiation were averaged over
1500 scans (about 16 minutes) during the first 400 minutes, 3000 scans (about
32 minutes) for the measurement at the 540\textsuperscript{th} minute, and 6000
scans afterward.

\section{Results and discussion}

\subsection{GPC results}

\begin{figure}
\begin{centering}
\includegraphics[scale=0.3]{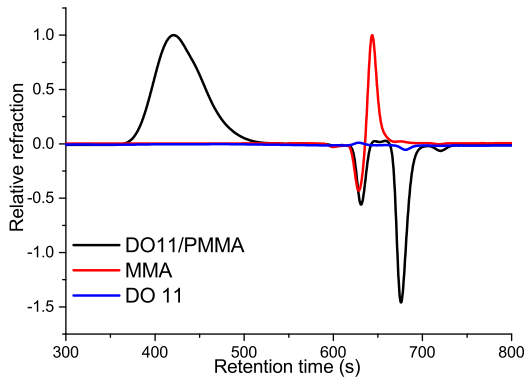}
\par\end{centering}
\protect\caption{\label{fig:RID}GPC results detected with RID. No peaks observed before 300 s.}
\end{figure}

\begin{figure}
\begin{centering}
\includegraphics[scale=0.3]{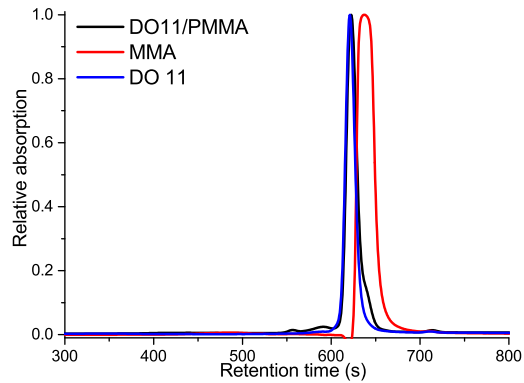}
\par\end{centering}
\protect\caption{\label{fig:SPD}GPC results detected with UV-Vis spectrometer at 254 nm. No peaks observed before 300 s.}
\end{figure}

The PTCR hypothesis suggests that DO11 (and 1AAQ) may undergo reversible
photodegradation with monomers. While DO11 dissolved
in dimethylformamide (DMF) was observed to undergo irreversible photodegradation
\cite{Howell2004}, DO11 dissolved in styrene and 1AAQ dissolved in
MMA have been observed to undergo the same reversible photodegradation as 
observed in dye-doped polymers \cite{Hung2015}
which supports the PTCR hypothesis. This indicates that DO11 and DMF do not form a 
metastable product as does DO11 and MMA or styrene after visible laser irradiation. 

Is the reversible photodegradation process observed in a dye-doped polymer due to the residual monomers, or are polymers involved as in the PTCR hypothesis? This hypothesis is tested by running pristine PMMA, DO11/PMMA, DO11 and MMA monomers through GPC.

MMA is visible in RID and UV-Vis spectrometry. PMMA is only visible in RID, while DO11 is only visible with UV-vis spectrometry. MMA is observed to peak at 644 sec when measured by RID in Figure \ref{fig:RID} and 637 sec when measured by UV-Vis spectrometry at 254 nm as shown in Figure \ref{fig:SPD}. Pristine DO11/PMMA shows no peak at 644 sec measured by RID indicating that no monomers are present in DO11/PMMA, which is confirmed by the UV-vis detector as no peak at 637 sec is observed. However, a peak at 623 sec suggests the presence of a small molecule. This peak overlaps with the signal of DO11. Since this molecule is invisible in RID - just as DO11 - this peaks can tentatively be ascribed to the dye. In RID, the peak at 421 sec arises from the polymer. Calibration towards PMMA (in sample DO11/PMMA) results in molar mass $\overline{M_{n}}$ of 9.2 kg/mol and dispersity {\DJ} of 2.4. Both measurements indicate that there are no MMA monomers in pristine DO11/PMMA sample. PMMA purchased from Aldrich is also found to contain no monomers from RID measurement. These results indicate that there exist no monomer residues in fresh DO11/PMMA that can contribute to the observation of reversible photodegradation.

\subsection{Computational results}

\begin{figure}[b!]
\begin{centering}
\includegraphics[scale=0.30]{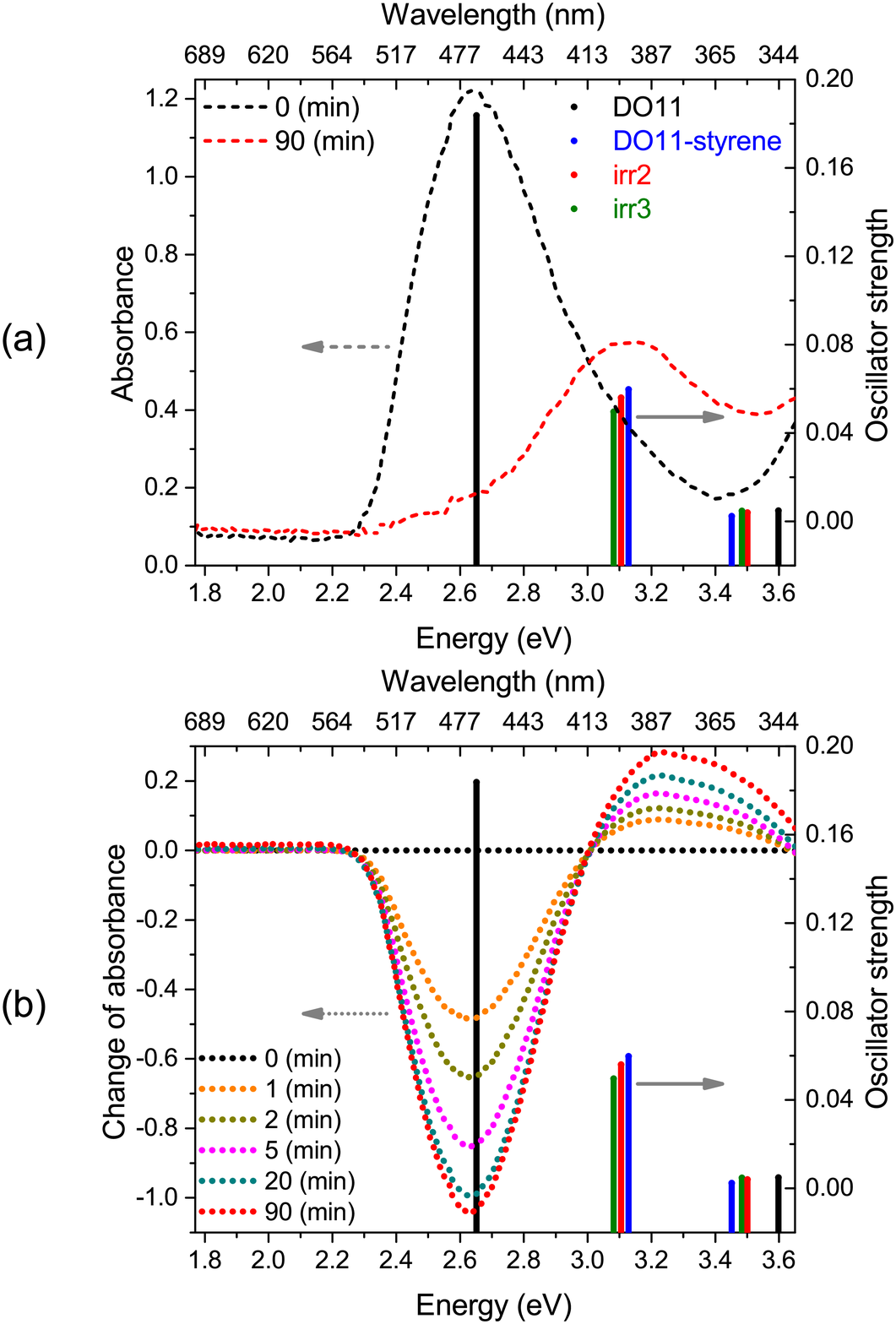}
\par\end{centering}

\protect\caption{\label{fig:130513ex3A9SAbsDFT2}(a) UV-Vis absorption spectra of
pristine DO11/PS and after 90 minutes of irradiation
are shown as dashed curves. The oscillator strengths of the proposed degraded
species are shown as vertical lines in proportion to their lengths. (b) Change of absorbance during decay
relative to the pristine sample (dotted curves) and the oscillator strengths
of the proposed degraded species (vertical lines).}
\end{figure}

\begin{figure}[b!]
\begin{centering}
\includegraphics[scale=0.30]{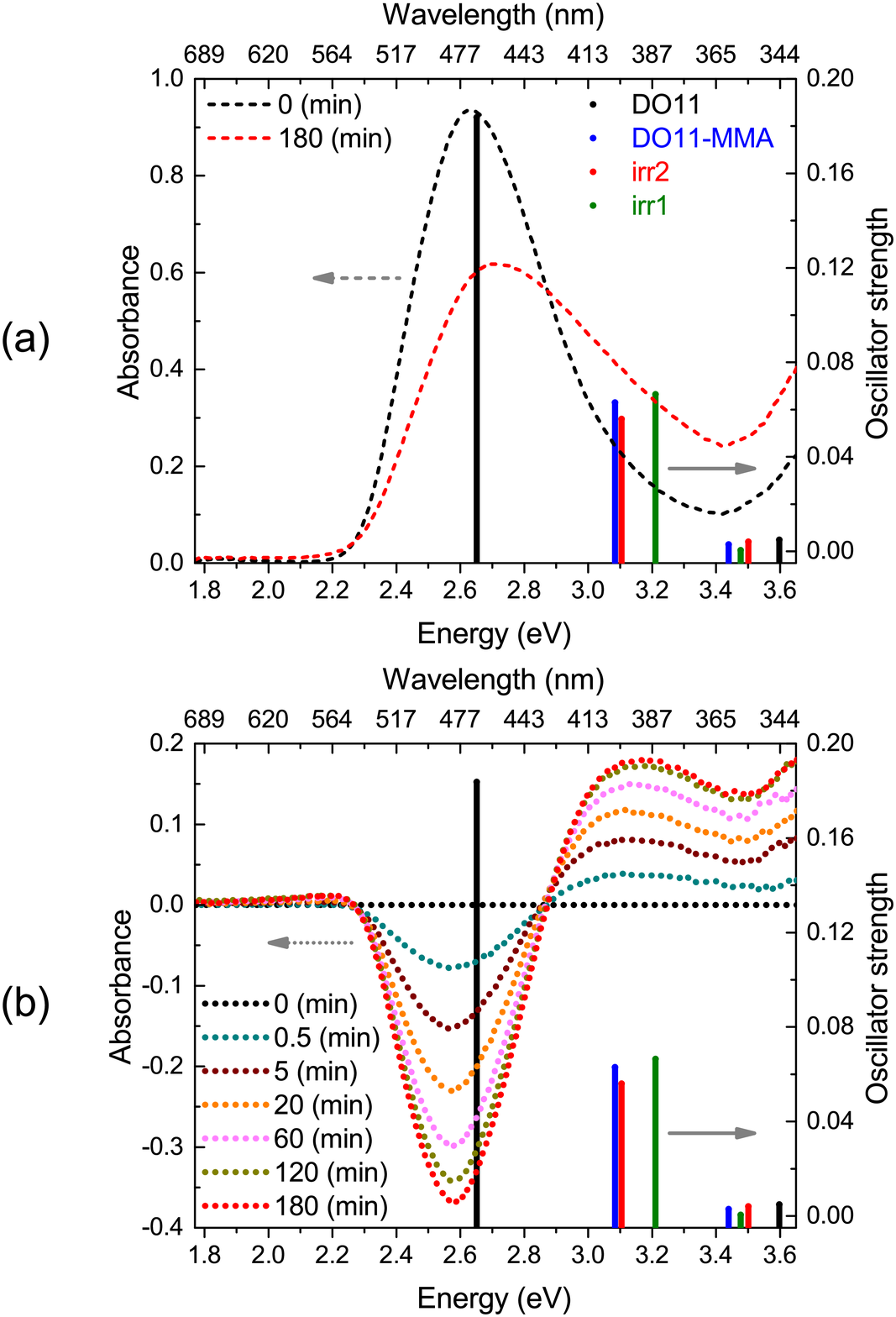}
\par\end{centering}

\protect\caption{\label{fig:A9MI35AbsDFT2}(a) UV-Vis absorption spectra of pristine
DO11/PMMA and after 180 minutes of irradiation are shown as dashed
curves. The oscillator strengths of the proposed degraded
species are shown as vertical lines in proportion to their lengths. (b) Change of absorbance during decay relative 
to the pristine sample (dotted curves) and the oscillator
strengths of the proposed degraded species (vertical lines).}
\end{figure}

The absorption spectrum of pristine dye doped in polymer can
be measured before the material is damaged and the absorption spectrum of all irreversibly-damaged
dye species together can be determined approximately from a measurement after the sample
has been irradiated for a long enough time such that the absorption spectrum no longer changes
with time. Figure \ref{fig:130513ex3A9SAbsDFT2}(a) and \ref{fig:A9MI35AbsDFT2}(a) show measurements before irradiation and after a long-time irradiation.
However, the absorption spectra of reversibly-damaged dye species
cannot be so easily determined due to the fact that they never appear alone and because 
the reversibly-damaged dye species recover making their contribution unknown.
Nonetheless, isosbestic points found in the change of absorption
spectra during photodegradation as shown in Figure \ref{fig:130513ex3A9SAbsDFT2}(b)
and \ref{fig:A9MI35AbsDFT2}(b) indicate that the absorption spectra
of the reversibly-damaged dye species are either similar to that of pristine
dye or of the irreversibly-damaged dye species in visible regime.

\begin{figure}[b!]
\begin{centering}
\includegraphics[scale=0.4]{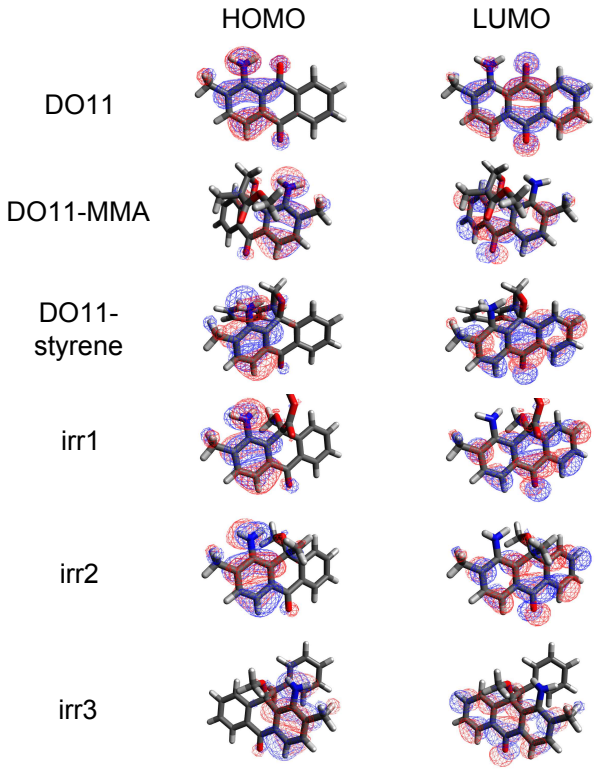}
\par\end{centering}

\protect\caption{\label{fig:HOMOLUMOPChem}HOMO and LUMO electron density of DO11 and
possible damaged species. Red and blue lobes indicate the electron
density with opposite phases in the electron wavefunctions. Colors
in molecular structures: black represents the carbon atom, gray
the hydrogen atom, red the oxygen atom and blue the nitrogen
atom.}
\end{figure}

The lowest electronic transition energies of DO11 and all degraded
species proposed in Figure \ref{fig:PChemSpecies} obtained using
TD-DFT calculations correspond to the transition between the highest
occupied molecular orbital (HOMO) and the lowest unoccupied molecular
orbital (LUMO). The electron density for HOMO and LUMO of each species
is shown in Figure \ref{fig:HOMOLUMOPChem}. The calculated electron density of HOMO-LUMO
transition for DO11 agrees with the computational results
for aminoanthraquinones in the literature \cite{Gordon1983,Inoue1972}:
the amine group is the principal electron donor of the ICT, and the
electron acceptors include both carbonyl groups, the center ring and
the unsubstituted ring. However, while the amine group remains the
principal electron donor of the ICT for all reversibly- and irreversibly-damaged
species, the carbonyl group adjacent to the amine group loses its
electron accepting ability even though the other electron acceptors are unchanged in
their electron accepting ability. Since the ICT is disrupted by the lowered
propensity for accepting electrons in the same way for all degraded species,
the absorption spectra of reversibly- and irreversibly-damaged species
are expected to be similar in visible regime.

Figure \ref{fig:130513ex3A9SAbsDFT2}(a) shows the absorption spectra of pristine and an approximation to irreversibly-damaged
DO11/PS. Figure \ref{fig:130513ex3A9SAbsDFT2}(b) shows the change of absorbance during photodegradation. The calculated oscillator
strengths and their corresponding excitation energies in visible regime
of pristine and reversibly- and irreversibly-damaged DO11 are also plotted
in Figure \ref{fig:130513ex3A9SAbsDFT2}. The agreement between the experimental and 
calculated results supports that the ICT is disrupted 
in the same way for both reversibly- and irreversibly-damaged species.
Figure \ref{fig:A9MI35AbsDFT2} shows that the 
DO11/PMMA data is not in as good agreement as the DO11/PS data 
possibly because the structure of toluene, the solvent chosen for
the TD-DFT calculation to approximate the polymer, is similar to a unit of PS but different from
a unit of PMMA. In addition, the calculated dipole moment of the damaged
species is larger than the pristine DO11 as listed in Table \ref{tab:DipoleMoment},
therefore the damaged species may interact more strongly with PMMA
than PS due to the methoxycarbonyl (\ce{COOCH3}) group in PMMA.
Nonetheless, the hypsochromic shift in the absorption spectrum and
the change in the magnitude of absorbance during photodegradation
qualitatively agree with the calculated oscillator strength and the
corresponding electronic transition energy for both DO11/PS and DO11/PMMA.

\begingroup
\squeezetable
\begin{table}[h]
\begin{centering}
\begin{tabular}{|c|c|c|c|c|c|c|}
\hline 
Molecules & DO11 & DO11-MMA & DO11-styrene & irr1 & irr2 & irr3\tabularnewline
\hline 
\hline 
Dipole moment (D) & 3.07 & 5.84 & 4.37 & 7.77 & 4.69 & 4.55\tabularnewline
\hline 
\end{tabular}
\par\end{centering}

\protect\caption{\label{tab:DipoleMoment}Dipole moment of DO11 and its photodegraded
species obtained using DFT calculation.}

\end{table}
\endgroup

\subsection{A simple model}

An alternative method to verify the above discussion is to calculate the absorbance of reversibly-photodegraded
dye species from experimental results by assuming that there is one
reversibly-damaged species and effectively one irreversibly-damaged
species. In this simple model, it is also assumed that damage
to the polymer does not contribute to the UV-Vis absorption
spectrum, i.e. it excludes the possibility of any absorption or light
scattering due to scattering centers generated from damaged polymer.
The calculated absorbance can be regarded as the result of the dominant
damage products or the average of several products with similar absorbance.

The absorbance, $A$, measured during decay and recovery at time $t$
can be expressed as
\begin{equation}
A\left(t\right)=n_{0}\left(t\right)A_{0}+n_{r}\left(t\right)A_{r}+n_{irr}\left(t\right)A_{irr},\label{eq:AbsTot}
\end{equation}
where $n_{0}$, $n_{r}$, $n_{irr}$, and $A_{0}$, $A_{r}$, $A_{irr}$
represent the fraction and the absorbance of the fresh, reversibly-damaged
and irreversibly-damaged dye species, respectively. Assuming that
dye molecules only convert between three species, $A_{0}$ and $A_{irr}$
can be experimentally determined from the absorption spectra of a fresh
sample and a sample that has been exposed to the pump laser for a
long enough time such that the absorption spectrum no longer changes.

The maximum fraction of irreversibly-damaged dye molecules $n_{irr}\left(t\rightarrow\infty\right)\equiv n_{irr}^{\infty}$
left in a partially recovered sample can be determined by the absorbance
at the time that recovery approaches completion,
\begin{eqnarray}
\nonumber
A\left(t\rightarrow\infty\right) &= & n_{0}\left(t\rightarrow\infty\right)A_{0}+n_{irr}\left(t\rightarrow\infty\right)A_{irr} \\
&= & A_{0}+n_{irr}^{\infty}\left(A_{irr}-A_{0}\right),
\label{eq:nIrr}
\end{eqnarray}
where we have used the assumption that the presence of only three
species yields $n_{0}\left(t\right)+n_{r}\left(t\right)+n_{irr}\left(t\right)=1$
with $n_{r}\left(t\rightarrow\infty\right)=0$. Thus, $n_{irr}^{\infty}$,
$A_{0}$, and $A_{irr}$ are obtained experimentally. The absorbance
of the reversibly-damaged species $A_{r}$, however, depends on how
the three species convert into each other and can be calculated by
rewriting Equation \ref{eq:AbsTot} at the time, $t_{0}$, that the
laser is turned off,
\begin{equation}
A_{r}=\frac{A\left(t_{0}\right)-n_{0}\left(t_{0}\right)A_{0}-n_{irr}\left(t_{0}\right)A_{irr}}{n_{r}\left(t_{0}\right)}.\label{eq:SigmaRe}
\end{equation}

\begin{figure}[b]
\begin{centering}
\includegraphics[scale=0.35]{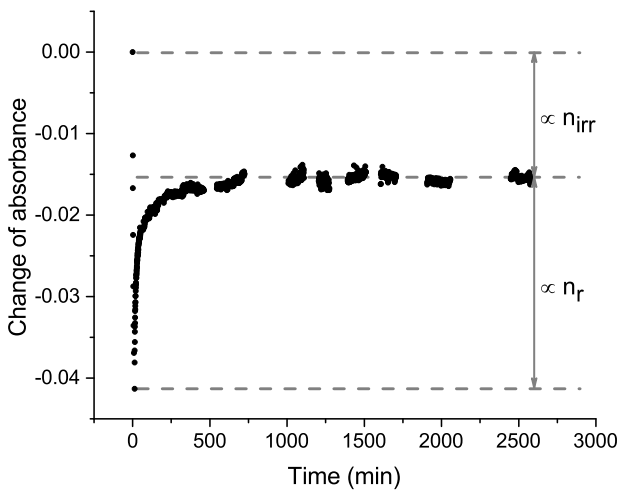}
\par\end{centering}
\protect\caption{\label{fig:140710ex1A9Mspecies}Change in absorbance of DO11/PMMA
during decay and recovery probed at 2.57 eV (482 nm). The sample was irradiated with
a 532 nm cw laser for 12 minutes then kept in the dark.}
\end{figure}

$n_{irr}$ is assumed to be a constant after the pump
laser is turned off, i.e. $n_{irr}\left(t\geq t_{0}\right)=n_{irr}^{\infty}$,
which implies that the reversibly-damaged species only recovers back
to the fresh dye but does not decay further into the irreversibly-damaged
species. With this assumption, $n_{r}\left(t_{0}\right)$ and $n_{irr}\left(t_{0}\right)$
are proportional to the change in absorbance between $t\rightarrow\infty$ and $t=t_{0}$ and between $t=0$ and $t\rightarrow\infty$,
as $n_{r}$ and $n_{irr}$ shown in Figure \ref{fig:140710ex1A9Mspecies}
and can be used to calculate $A_{r}$ from Equation \ref{eq:SigmaRe}. This
assumption can be relaxed to $n_{irr}(t\geq t_{0})\leq n_{irr}^{\infty}$,
in which case the reversibly-damaged species may either decay to the
irreversibly-damaged species or recover to the pristine dye, and $n_{r}\left(t_{0}\right)$
and $n_{irr}\left(t_{0}\right)$ can be numerically adjusted to generate
$A_{r}$ using Equation \ref{eq:SigmaRe}. Note that if this is so, the total fraction
of damaged molecules after the laser is turned off at time $t_{0}$,
$n_{r}(t\geq t_{0})+n_{irr}(t\geq t_{0})$, is not
necessarily the same as in the former case where irreversibly-damaged species is limited to conversion to the pristine molecule. 


\begin{figure}[b]
\begin{centering}
\includegraphics[scale=0.30]{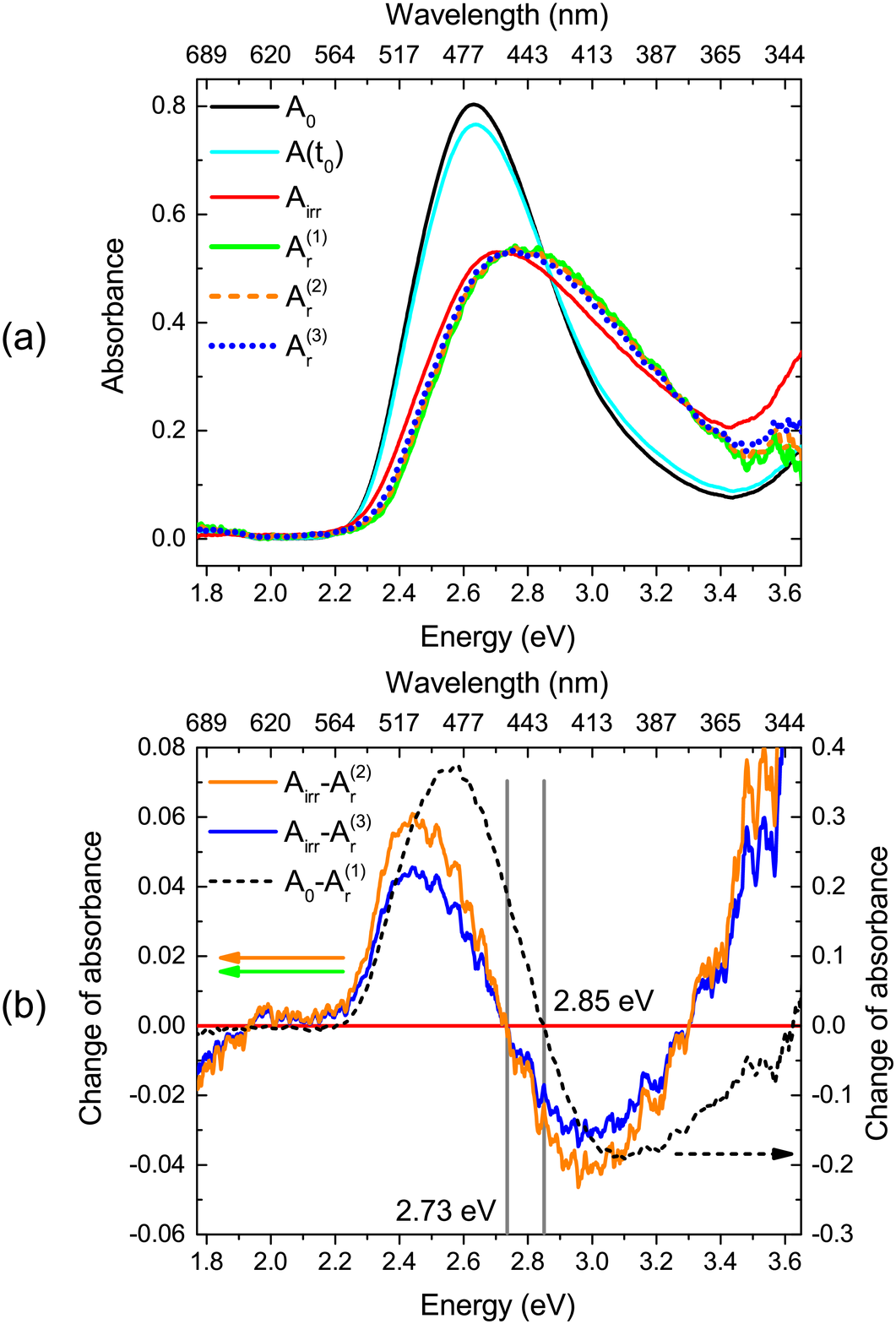}
\par\end{centering}
\protect\caption{\label{fig:140710ex1A9MI35AbsSpecies}(a) Absorbance of pristine
DO11/PMMA ($A_{0}$), reversibly-damaged DO11/PMMA ($A_{r}^{(1)}$,
$A_{r}^{(2)}$ and $A_{r}^{(3)}$, depending on the conditions 
described in the text) and the irreversibly-damaged DO11/PMMA
($A_{irr}$). $A\left(t_{0}\right)$ is the absorbance of the DO11/PMMA
sample measured when the pump laser was
turned off. (b) Difference of absorbance between the irreversibly-damaged
species and the reversibly-damaged species ($A_{r}^{(2)}$ and $A_{r}^{(3)}$,
see text and Table \ref{tab:nFracA9M}), and between the pristine
dye and the reversibly-damaged species $A_{r}^{(1)}$.}
\end{figure}

\begingroup
\begin{table*}
\begin{centering}
\begin{tabular}{|c|c|c|c|}
\hline 
 & $n_{0}(t_{0})$ ($\times 10^{-2}$) & $n_{r}(t_{0})$ ($\times 10^{-2}$) & $n_{irr}(t_{0})$ ($\times 10^{-2}$) \tabularnewline
\hline 
\hline 
$A_{r}^{(1)}$ & 88.26$\pm$0.03 & 7.35$\pm$0.03 & 4.40$\pm$0.03 ($n_{irr}(t\geq t_{0})=n_{irr}^{\infty}$) \tabularnewline
\hline 
$A_{r}^{(2)}$ & 88.26 & 8.79 & 2.95\tabularnewline
\hline 
$A_{r}^{(3)}$ & 88.26 & 11.74 & 0\tabularnewline
\hline 
\end{tabular}
\par\end{centering}

\protect\caption{\label{tab:nFracA9M}Fractions of the three species $n_{0}$, $n_{r}$
and $n_{irr}$ at time $t_{0}$ that result in the corresponding absorbance
of the reversibly-damaged DO11/PMMA species $A_{r}^{(1)}$, $A_{r}^{(2)}$
and $A_{r}^{(3)}$ shown in Figure \ref{fig:140710ex1A9MI35AbsSpecies}(a).}
\end{table*}
\endgroup

In practice, $A_{r}$ and $A_{irr}$ are determined from two independent
experiments, since the reversibly-damaged species recovers
but irreversibly-damaged ones don't. 
The experimental uncertainty thus has contributions from noise from
the white light source, the spectrometer and uncertainties due to the determination of the spectra of the two damaged species. However, absorption
spectra of the damaged species can be determined qualitatively.

A DO11/PMMA thin film of concentration 9 g/L is subjected to long-time
photodegradation, and the absorbance of the pristine
and irreversibly-damaged sample are shown in Figure \ref{fig:A9MI35AbsDFT2}(a).
Another DO11/PMMA sample of the same concentration is characterized
during recovery after irradiated by a 532 nm
cw laser with a peak intensity of 2.09 W/cm\textsuperscript{2} for 12
minutes. The absorbance of the pristine DO11/PMMA sample,
$A_{0}$, and the absorbance of the DO11/PMMA sample
measured at the 12\textsuperscript{th} minute, $A\left(t_{0}\right)$, where $t_{0}=12$ min, are shown
in Figure \ref{fig:140710ex1A9MI35AbsSpecies}(a). Since both samples
have the same concentration, $A_{irr}$ of the later sample can be
obtained using the Lambert-Beer law and is plotted in Figure \ref{fig:140710ex1A9MI35AbsSpecies}(a).
The absorption spectra between 2520 and 2550 minutes were averaged
as an approximation of $A(t\rightarrow\infty)$ in Equation \ref{eq:nIrr},
and $A(t\rightarrow\infty)$ between 477 nm and 487 nm were used to
determine the averaged ratio of $n_{1}/n_{2}$. Using Equation \ref{eq:nIrr},
$A_{0}$ and $A_{irr}$ in Figure \ref{fig:140710ex1A9MI35AbsSpecies}(a),
$n_{irr}$ is determined and also averaged between 477 nm and 487
nm. Thus, $A_{r}$ is determined using Equation \ref{eq:SigmaRe}
with the assumption that $n_{irr}(t\geq t_{0})$ is constant, as 
shown in Figure \ref{fig:140710ex1A9MI35AbsSpecies}(a) by $A_{r}^{(1)}$.

The assumption $n_{irr}(t\geq t_{0})=n_{irr}^{\infty}$ can be relaxed
to $n_{irr}(t_{0})\leq n_{irr}^{\infty}$ to allow the reversibly-damaged
species to either decompose to the irreversibly-damaged species or to recover
to the fresh dye. Also assumed is that the total fraction of damaged molecules
after time $t_{0}$, $n_{r}(t\geq t_{0})+n_{irr}(t\geq t_{0})$, is
the same as previously calculated, which may not be true but is used
as a guess to study $A_{r}$. Two other values of $n_{irr}(t_{0})$
have been chosen to calculate $A_{r}$, and parameters $n_{0}$, $n_{r}$
and $n_{irr}$ at time $t_{0}$ are tabulated in Table \ref{tab:nFracA9M}
and the calculated absorbance of the reversibly-damaged species, $A_{r}^{(2)}$
and $A_{r}^{(3)}$, are plotted in Figure \ref{fig:140710ex1A9MI35AbsSpecies}(b). Note that the small uncertainty in Table \ref{tab:nFracA9M} results from the analysis done using the peak of changing absorbance (between 477 nm and 487 nm). The purpose of this analysis is to qualitatively determine the absorption spectrum of reversibly-damaged species based on the simple three species model, so the small uncertainty of population does not necessarily mean that the population is accurately determined.

The large difference between the absorbance of the reversibly-damaged
species and $A_{irr}$ observed in the UV regime indicates that the
absorbance of reversibly- and irreversibly-damaged dye species
is more distinguishable there than in the visible region. However,
this statement is based on the assumption that there is only one reversibly-
and (effectively) one irreversibly-damaged dye species generated during
photodegradation and that the damaged PMMA does not contribute to the absorption
spectrum.

\begin{figure}[b]
\begin{centering}
\includegraphics[scale=0.35]{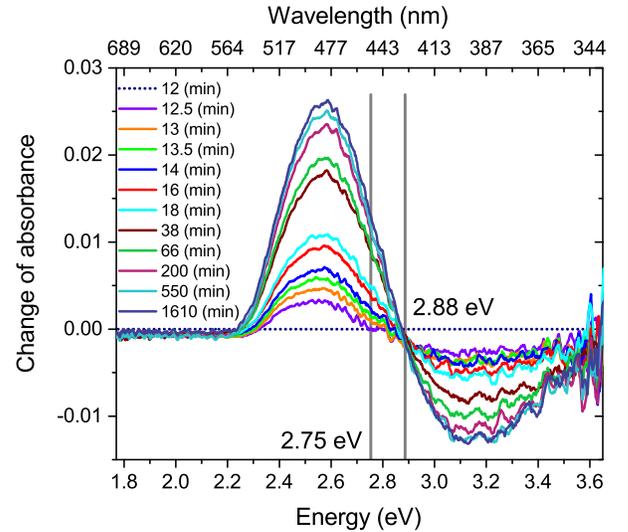}
\par\end{centering}
\protect\caption{\label{fig:140710ex1A9MAbsRe}Change in absorbance of DO11/PMMA thin
film relative to the spectrum taken at the 12\protect\textsuperscript{th}
minute, the time at which the pump beam is turned off.}
\end{figure}

The absorbance of the reversibly- and irreversibly-damaged species, $A_{r}^{(1)}$ and $A_{irr}$, are similar in the visible regime, which
is consistent with the results of TD-DFT calculations. The amplitude
of the absorbance difference between the fresh molecule, $A_{0}$, and the reversibly-damaged species, $A_{r}^{(1)}$, is
approximately 5 times larger than that between the irreversibly-damaged species, $A_{irr}$, and the reversibly-damaged species, $A_{r}^{(2)}$,
and the isosbestic point is found to be at 2.85 eV for recovery
and at 2.73 eV for further decay from the reversibly-damaged species as shown
in Figure \ref{fig:140710ex1A9MI35AbsSpecies}(b). The change of absorbance
during recovery is plotted with isosbestic points indicated for the
spectrum taken at the 12.5\textsuperscript{th} minute and at later
times in Figure \ref{fig:140710ex1A9MAbsRe} (recovery starts at the 12\textsuperscript{th} minute). The result indicates
that some reversibly-damaged dye species further decay into irreversibly-damaged
species after the pump beam is turned off. It also suggests that the
further decay of the reversibly-damaged species has a faster rate than the recovery rate since the isosbestic
point is observed at 2.75 eV in the beginning of the 
process (after the pump beam is turned off). However, the isosbestic point quickly shifts to 2.88 eV suggesting
that the recovery process dominates when the damaged sample is allowed to stay in the dark, which is verified by damaging
a DO11/PMMA thin film with a cw laser and probing the recovery using both
UV-Vis absorption spectroscopy and amplified spontaneous emission
that only comes from pristine DO11 \cite{Hung2015}.

The absorption spectrum of the reversibly-damaged dye
species in DO11/PMMA calculated using the above model qualitatively
agrees with the one obtained from TD-DFT calculations.
The same agreement is found in DO11/PS thin film samples \cite{Hung2015}. Unlike DO11/PMMA, it is
concluded that most of the reversibly-damaged dye molecules further decay to the irreversibly-damaged
species, which is in agreement with the observed
small amount of recovery in a typical DO11/PS sample \cite{Hung2015}.

\subsection{FTIR results}

In the PTCR hypothesis, the polymer is involved in the process of 
photodegradation of dye and may also recover as dye recovers. 
A UV-Vis absorption spectrum shows peaks due to the dyes in a dye-doped
polymer; but since PS and PMMA are transparent in this wavelength
range, the contribution from the host polymer is negligible. However,
infrared (IR) absorption spectroscopy allows the study of both dye
and polymer, so we apply FTIR spectroscopy
to monitor the change of chemical bonds in the dye molecules and in
the host polymer in reversible photodegradation.

1AAQ/PMMA is chosen for FTIR characterization studies because of the
abundant literature on 1AAQ. PS is not studied in the FTIR experiment
due to the small degree of recovery, which is difficult to
distinguish from noise. A polymerized sample sandwiched between two
glass substrates cannot be used in FTIR experiments because glass
absorbs IR radiation in the region of interest. So, silicon wafers, 
which provide a good IR window for this study, are used as
substrates for spin-coated films.

A typical polymerized DO11/PMMA sample used in this study has a concentration
of 9 g/L, which corresponds to about 0.32\% number ratio of DO11 in
MMA monomer. The same number ratio of 1AAQ in MMA monomer corresponds
to 8.5 g/L of 1AAQ/PMMA sample. IR absorption from the
dye molecules is much weaker than from PMMA at typical dye concentrations.
In order to observe possible changes in the IR spectrum from both
dye and polymer, the concentration of 1AAQ/PMMA is increased to 105
g/L, about a 3.96\% number ratio of 1AAQ in MMA monomer, resulting in
a dye concentration of more than 12 times of a typical polymerized
sample. 

Even though samples of 1AAQ/PMMA are prepared differently,
the observed reversible 
photodegradation in spin-coated samples is the same as in polymerized 1AAQ/PMMA \cite{Hung2015}. 
Though the FTIR experiment is performed in vacuum, 
and the UV-Vis spectroscopy experiment is open to the air with sandwiched samples  
between two glass substrates, a test on reversible photodegradation of spin-coated samples 
suggests that the effect of air is likely negligible during photodegradation 
and recovery as described in Section A of supplemental material.

\begin{figure*}[h]
\begin{centering}
\includegraphics[scale=0.5]{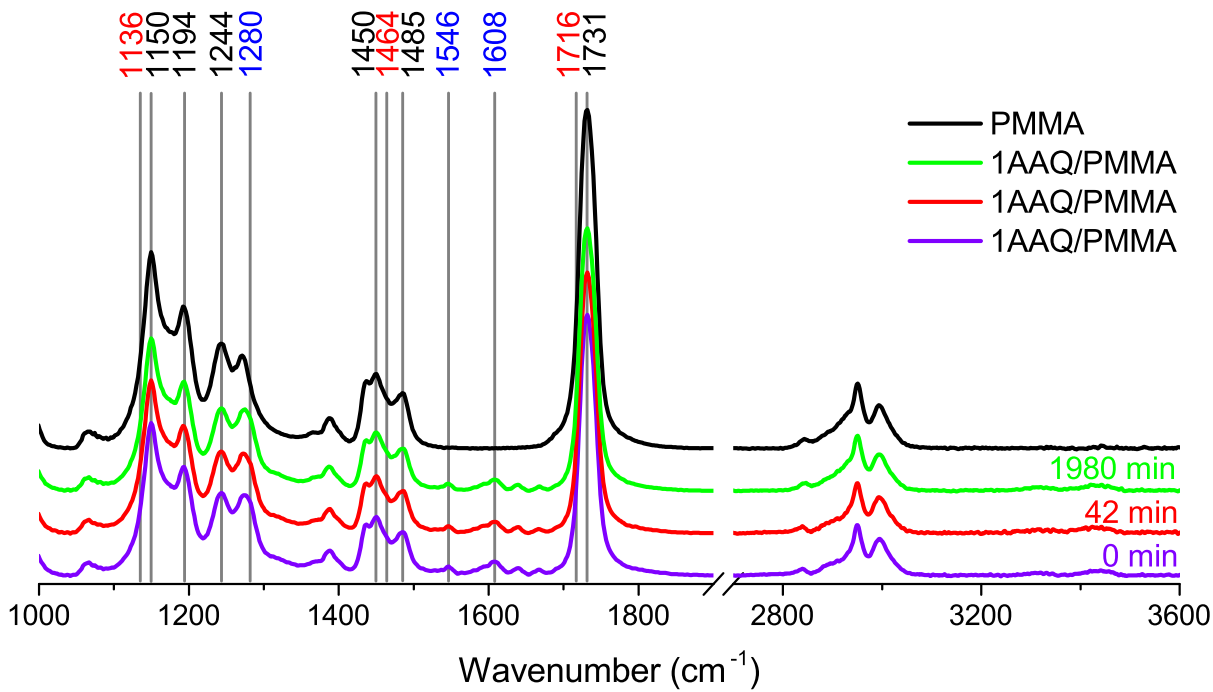}
\par\end{centering}

\protect\caption{\label{fig:140729ex2Y105MreIRspec}IR absorbance of PMMA and 1AAQ/PMMA
at 0, 42 and 1980 min after the silicon substrate contribution is subtracted. IR absorption peaks that vary during recovery
between 1000 and 1800 cm\protect\textsuperscript{-1} are labeled
in black for PMMA, blue for 1AAQ and red for absorption peaks that
do not belong to PMMA and 1AAQ.}
\end{figure*}

\begin{figure*}[h]
\begin{centering}
\includegraphics[scale=0.5]{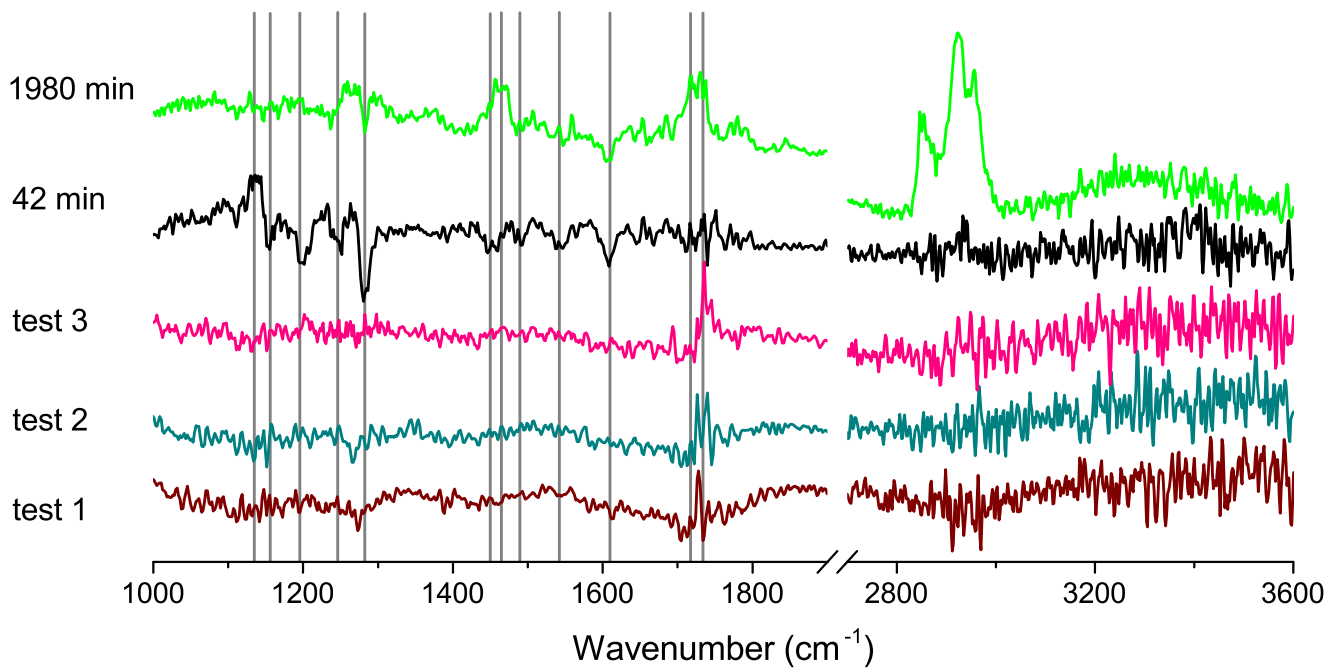}
\par\end{centering}

\protect\caption{\label{fig:140729ex2Y105MftirSNratio}Comparison of the change in the spectrum of a fresh sample over time (test 1, test 2 and test 3) as a baseline and the
change in IR absorbance of 1AAQ/PMMA during recovery at 42 minutes and 1980 minutes. The 42\textsuperscript{nd} minute and 1980\textsuperscript{th} 
minute are respectively the first spectrum and the last one acquired after 30 minutes irradiation relative 
to the spectrum of the pristine sample. The vertical lines correspond to the 
varying IR absorption bands during recovery as assigned in Figure \ref{fig:140729ex2Y105MftirPTCR}.}
\end{figure*}

The IR spectra of 1AAQ/PMMA
as a function of time are obtained by subtracting the spectrum of each sample
from that of the silicon wafer, and some results of which are shown in Figure \ref{fig:140729ex2Y105MreIRspec} together with the IR absorption spectrum of PMMA.
The spectrum of 1AAQ/PMMA at the 42\textsuperscript{nd} minute is the first spectrum
acquired after 30 minutes of irradiation. The spectrum of 1AAQ/PMMA at the 1980\textsuperscript{th} minute 
is the last spectrum taken during recovery.

The 1AAQ/PMMA sample needs to be rotated $90{}^{\circ}$
for laser irradiation and rotated back to its original position to
record IR spectra during recovery, which introduces additional
variability and is studied by simulating the
experiment without irradiating the sample as follows. An IR absorption
spectrum of 1AAQ/PMMA is  obtained by averaging 6000 scans. The sample
is then rotated $90{}^{\circ}$ (without laser irradiation) and rotated
back to the original position. Subsequently 1000 scans are averaged
to obtain another spectrum and the procedure is repeated. 45 minutes after
taking the second spectrum, another spectrum 
is similarly acquired. The change in the IR absorption spectrum is obtained by subtracting
the first spectrum, taken before the sample is rotated, from the later
ones. The results of two full cycles of rotation are plotted as ``test
1'' and ``test 2'', and the one taken 45 minutes after ``test
2'' is plotted as ``test 3'' in Figure \ref{fig:140729ex2Y105MftirSNratio}. 
The change in first and the last spectra during recovery of a degraded sample relative to
the spectrum taken before photodegradation are also plotted in Figure \ref{fig:140729ex2Y105MftirSNratio} and
are found to be distinguishable from noise, except for peaks near
1731 cm\textsuperscript{-1} due to the nearly saturated absorbance
at 1731 cm\textsuperscript{-1}.

\begingroup
\begin{table}
\begin{centering}
\begin{tabular}{|>{\centering}p{1.5cm}| >{\centering}p{2.0cm}|>{\centering}p{3cm}|}
\hline 
Wavenumber (cm\textsuperscript{-1}) & Molecule & Corresponding vibration\tabularnewline
\hline 
\hline 
\multicolumn{1}{|c|}{{\color{gCCOC} 1150}, {\color{gCCOC} 1194}, {\color{gCCOC}
1244}} & \multirow{4}{*}{\includegraphics[scale=0.4]{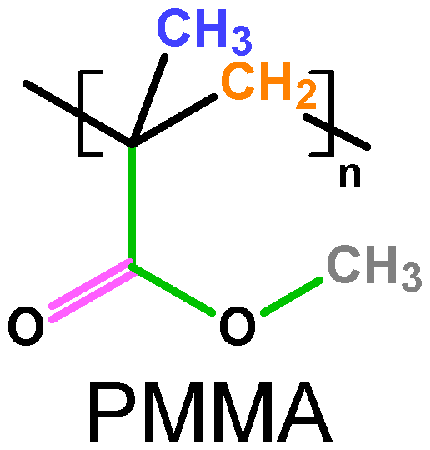}} & \ce{C-C-O-C} stretching\tabularnewline
\cline{1-1} \cline{3-3} 
{\color{oCH2} 1450} &  & \ce{CH2}, $\alpha$-\ce{CH3} and \ce{(O)CH3} bending\tabularnewline
\cline{1-1} \cline{3-3} 
{\color{bCH3} 1485} &  & $\alpha$-\ce{CH3} asymmetric deformation\tabularnewline
\cline{1-1} \cline{3-3} 
{\color{rCO} 1731} &  & \ce{C=O} stretching\tabularnewline
\hline 
{\color{Cyan} 1280} & \multirow{4}{*}{\includegraphics[scale=0.3]{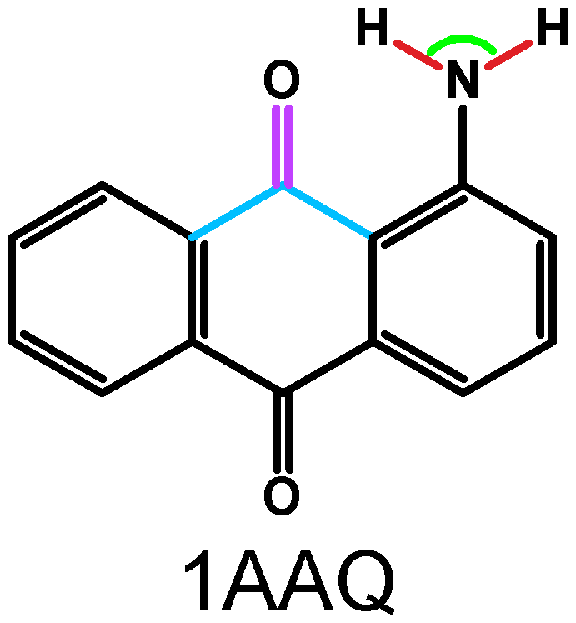}} &  \ce{C-C-C} stretch of ketones\tabularnewline
\cline{1-1} \cline{3-3} 
{\color{gNH2} 1546} &  & in-plane \ce{NH2} scissoring\tabularnewline
\cline{1-1} \cline{3-3} 
{\color{vCO} 1608} &  & \ce{C=O} stretching\tabularnewline
\cline{1-1} \cline{3-3} 
{\color{rNH2} 3320}, {\color{rNH2} 3440} &  & \ce{N-H} stretching\tabularnewline
\hline 
1136 & \includegraphics[scale=0.3]{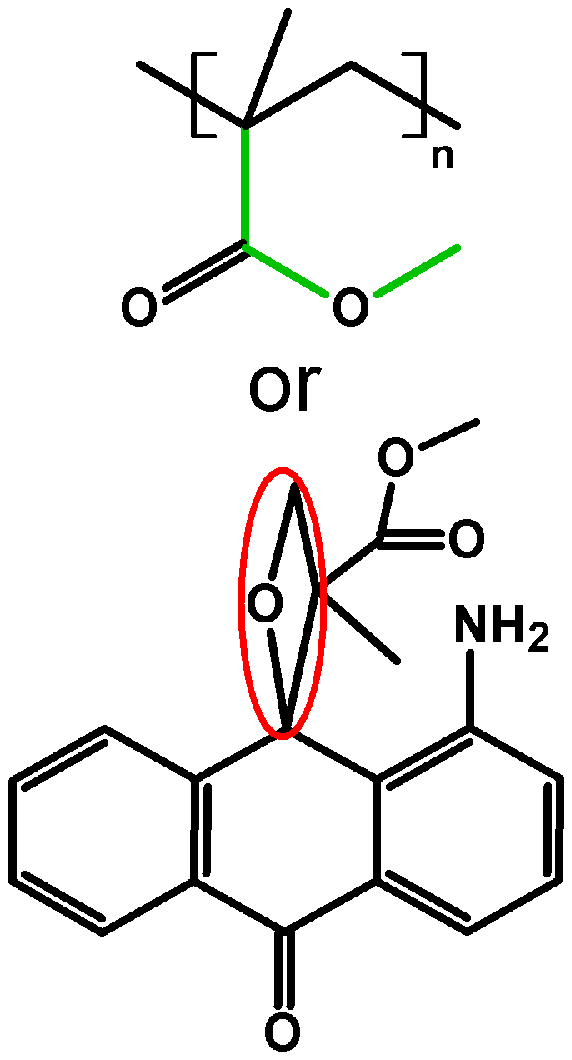} & See text\tabularnewline
\hline 
{\color{rsp3} 1464} & \includegraphics[scale=0.4]{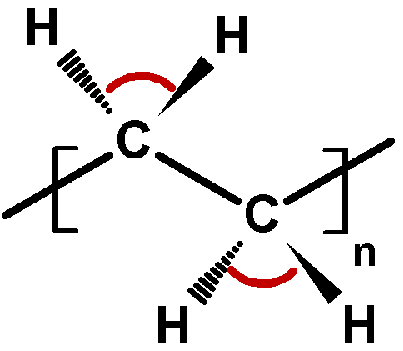} & \ce{CH2} scissoring\tabularnewline
\hline 
{\color{yCO} 1716} & \includegraphics[scale=0.4]{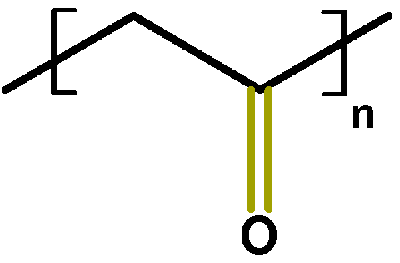} & \ce{C=O} stretching\tabularnewline
\hline 
{\color{Salmon} 2850-3000} & \includegraphics[scale=0.4]{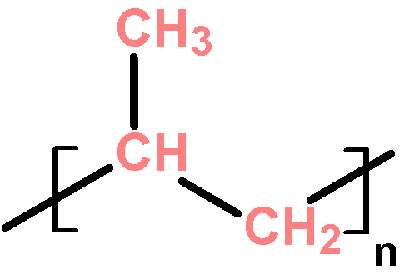} & sp\textsuperscript{3} \ce{CH} stretching\tabularnewline
\hline 
\end{tabular}
\par\end{centering}

\protect\caption{\label{tab:TabIRpeaks}The assignment of IR absorption bands with
each assignment color-coded according to the corresponding molecular structure.
The molecular structures shown for molecules other than PMMA and 1AAQ
are illustrations of the corresponding IR bands, not necessarily
the actual molecules. Details of the assignment of the IR bands are
described in the text.}
\end{table}
\endgroup

\begin{figure*}
\begin{centering}
\includegraphics[scale=0.46]{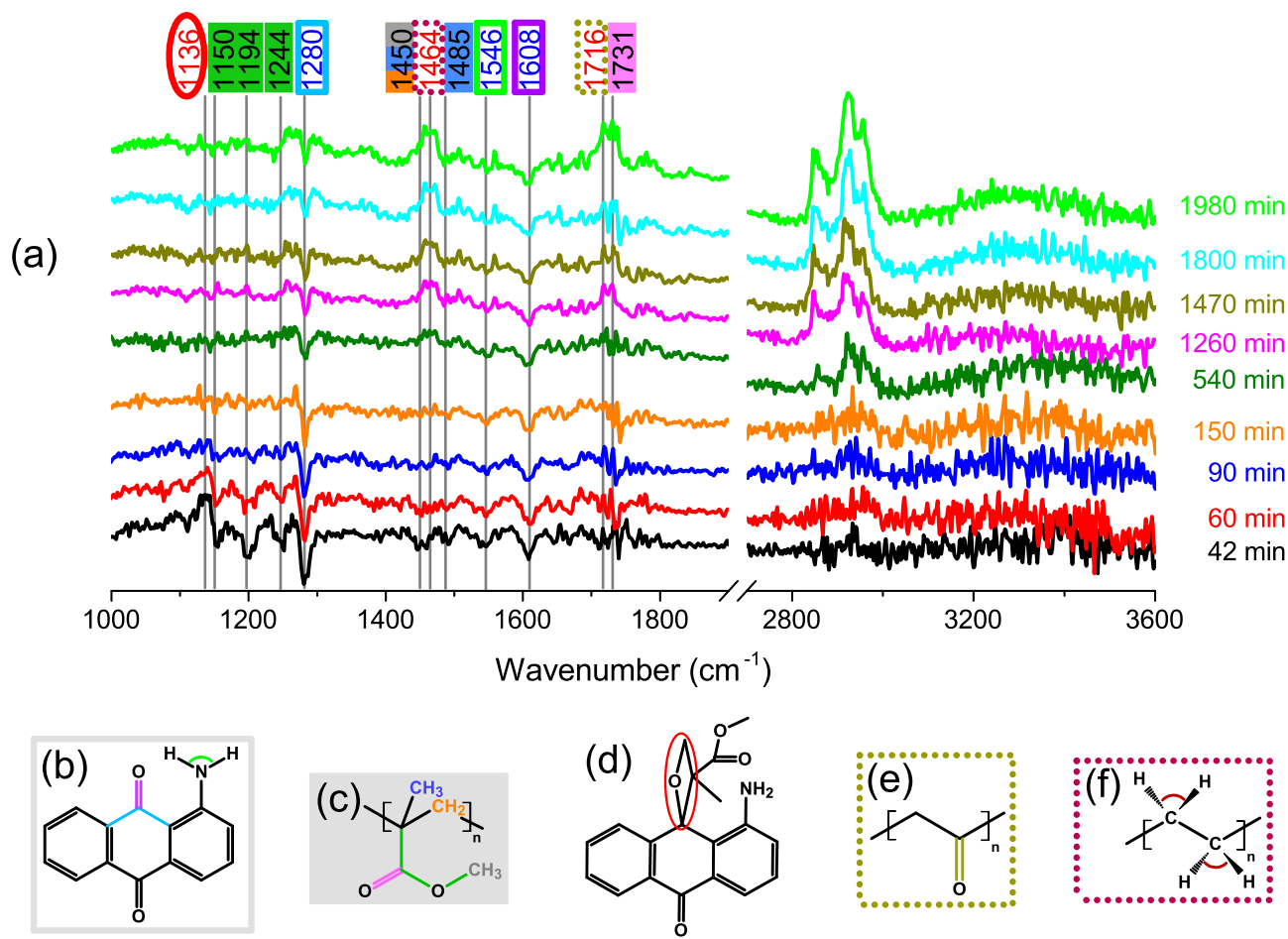}
\par\end{centering}

\protect\caption{\label{fig:140729ex2Y105MftirPTCR}(a) Evolution of the change in the
IR spectrum relative to pristine 1AAQ/PMMA after 30 minutes of
irradiation and recovery thereafter. The boxed wavenumbers are color-coded in terms of the bonds indicated in the structures below for (b) 1AAQ shown with hollow solid boxes, (c) PMMA shown with filled boxes, (d) 1AAQ-MMA oxetane shown red circle, (e) and (f) other structures
originate from the background of the system shown in dotted boxes.}
\end{figure*}

The change in IR absorbance of 1AAQ/PMMA during recovery relative to the spectrum 
of the pristine sample is shown in Figure \ref{fig:140729ex2Y105MftirPTCR}(a). 
IR absorption peaks that change over time 
are assigned to the corresponding molecular vibration modes as described
below and tabulated in Table \ref{tab:TabIRpeaks}. According to Nagai
et al., the bands at 1150 and 1194 cm\textsuperscript{-1} are associated
with coupled vibrations of skeletal stretching and internal \ce{C-H}
deformation modes, and the band at 1244 cm\textsuperscript{-1} is
associated with the coupled \ce{C-C-O} and \ce{C-O} stretching
vibrations \cite{Nagai1962,Nagai1963}. However, the bands between
1150 and 1300 cm\textsuperscript{-1} have been attributed later to
strong coupling between stretching vibrations of \ce{C-C-O} and
\ce{C-O} or \ce{C-O-C} and \ce{C-O} of the ester group
(\ce{R1-COO-R2}) \cite{Willis1969,Silverstein1998b,Workman2001}.
The more current assignment is adopted for the bands at 1150, 1194
and 1244 cm\textsuperscript{-1} to be coupled \ce{C-C-O-C} stretching
vibrations. The bending vibrations of \ce{CH2}, $\alpha$-\ce{CH3}
(in which $\alpha$ represents the first carbon atom that is attached
to the functional group \ce{COOCH3}) and \ce{(O)CH3} strongly
overlap around the band at 1450 cm\textsuperscript{-1} \cite{Nagai1962,Nagai1963,Schneider1979}.
The peak at 1485 cm\textsuperscript{-1} can be assigned to the $\alpha$-\ce{CH3}
asymmetric deformation vibration of PMMA \cite{Nagai1962,Nagai1963,Schneider1979}.
While the peak at 1136 cm\textsuperscript{-1} may be a new peak,
it might also be the result of a frequency shift
of the 1150 cm\textsuperscript{-1} peak under irradiation that recovers
after the laser is turned off.

The peak at 1280 cm\textsuperscript{-1} is of the same magnitude
as the 1244 cm\textsuperscript{-1} peak in the IR spectrum of pristine
1AAQ/PMMA in Figure \ref{fig:140729ex2Y105MreIRspec}, but the 1280 cm\textsuperscript{-1}
is located at the shoulder of the IR absorption peak of PMMA at 1270
cm\textsuperscript{-1}, which is smaller than the 1244 cm\textsuperscript{-1}
peak from pure PMMA, so the peak at 1280 cm\textsuperscript{-1}
is due to 1AAQ and can be attributed to \ce{C-C-C} stretch of
ketones, i.e. \ce{C-C(=O)-C} \cite{Silverstein1998b}. The peak
at 1546 cm\textsuperscript{-1} can be assigned to the in-plane \ce{NH2}
scissoring of 1AAQ, though is 34 cm\textsuperscript{-1} lower wavenumbers than
the usual frequency range in organic compounds \cite{Silverstein1998b},
which could be caused by the intramolecular hydrogen bond. The band
at 1608 cm\textsuperscript{-1} can be attributed to the vibration
of \ce{C=O} adjacent to the amine group \cite{Flett1948} of
1AAQ. It is worth noticing that there is no change in the band at
1667 cm\textsuperscript{-1}, which can be assigned to the vibration
of the other \ce{C=O} group of the 1AAQ molecule \cite{Flett1948}
and is observed in Figure \ref{fig:140729ex2Y105MreIRspec} (therefore
not labeled in the figure). The symmetric and asymmetric \ce{N-H}
stretching modes of the amine group at about 3320 cm\textsuperscript{-1}
and 3440 cm\textsuperscript{-1} \cite{Flett1948} are too noisy to
determine whether they changed after irradiation.

Some of the observed changing IR absorption peaks are not due to PMMA
and 1AAQ and originate from changes in the background of the system, such as contamination from pump oil.
The peak at 1464 cm\textsuperscript{-1} is known to be the \ce{CH2}
scissoring band of hydrocarbons \cite{Silverstein1998b,LinVien1991}.
The band at 1716 cm\textsuperscript{-1} originates from the \ce{C=O}
stretching vibration \cite{Silverstein1998b} but not from the
aminoanthraquinones, which possess a \ce{C=O} stretch frequency
of 1610$\sim$1680 cm\textsuperscript{-1} \cite{Flett1948}, nor from
PMMA, which exhibits a \ce{C=O} stretch near 1731 cm\textsuperscript{-1}
\cite{Fowkes1984}. The change of IR absorption bands between 2850
and 3000 cm\textsuperscript{-1} are known to be the sp\textsuperscript{3}
\ce{CH} stretching vibration \cite{LinVien1991,Silverstein1998b}.
These peaks originate from the FTIR system as described in Section B of supplemental material.

In the PTCR hypothesis, it is posited that 1AAQ undergoes photocycloaddition
to MMA that is generated from thermally degraded PMMA to form an oxetane
structure - the reversibly-damaged species, or undergoes other (photo)chemical
reactions with fragments thermally degraded from PMMA to form the
irreversibly-damaged species. Figure \ref{fig:140729ex2Y105MftirPTCR}(b)$\sim$(f) 
illustrate the possible damage to 1AAQ and PMMA
and the corresponding change in IR absorption peaks.

The reduction of \ce{C-C-O-C} stretching vibrations of PMMA (1150$\sim$1300
cm\textsuperscript{-1}) after irradiation indicates the scission
of the methoxycarbonyl group (\ce{COOCH3}) from PMMA under irradiation.
The decrease of \ce{CH2} bending (1450 cm\textsuperscript{-1})
and $\alpha$-\ce{CH3} asymmetric deformation (1485 cm\textsuperscript{-1})
of PMMA after irradiation suggests that depolymerization and dissociation
of polymer backbones take place under irradiation. Photocycloaddition
between 1AAQ and the depolymerized MMA causes the decrease of the
\ce{C=O} vibration peak (1608 cm\textsuperscript{-1}) and the
\ce{C-C-C} stretch peak of ketone (1280 cm\textsuperscript{-1})
in 1AAQ. 1AAQ and other radicals and fragments dissociated from PMMA
can also undergo (photo)chemical reaction to form irreversibly-damaged
species and cause the same change in the IR peaks. Note that there
is no change from the other \ce{C=O} group of the 1AAQ molecule
at 1667 cm\textsuperscript{-1} \cite{Flett1948},
which is consistent with the conclusion obtained from the linear absorption
spectroscopy measurements and the TD-DFT calculations in DO11. The
reduction in IR peaks of PMMA after irradiation provides evidence
of polymer degradation, which supports the PTCR hypothesis.

\begin{figure}
\begin{centering}
\includegraphics[scale=0.4]{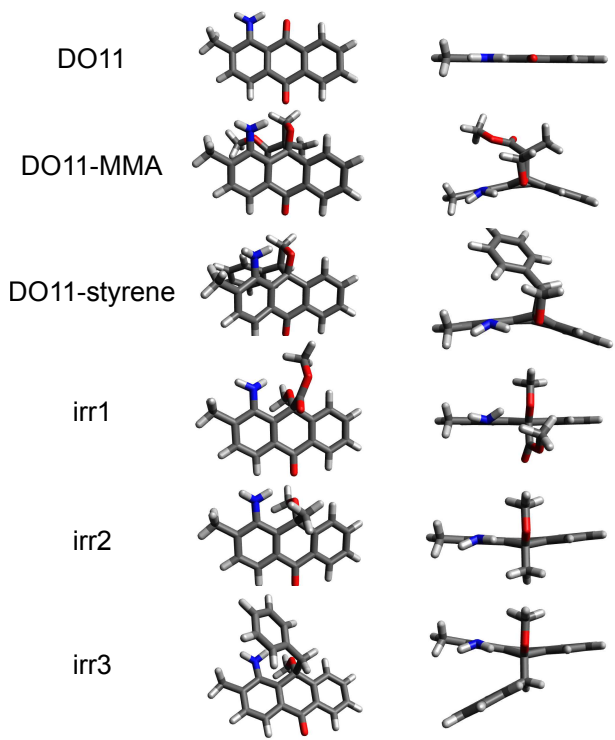}
\par\end{centering}

\protect\caption{\label{fig:DFTsturctures}Molecular structures of DO11 and possible
damaged species. Colors in molecular structures have the same meaning
as in Figure \ref{fig:HOMOLUMOPChem}.}
\end{figure}

The in-plane \ce{NH2} scissoring of 1AAQ at 1546 cm\textsuperscript{-1}
may be weakened after irradiation, though the change of the IR peak is
noisy. It may be attributed to similar behaviors observed from the
ground state geometry optimization of DO11 and its damaged-species
in the DFT calculations, i.e. the amine group in the reversibly- and
irreversibly-damaged species, though still fairly planar, is observed
to become slightly out-of-plane or slightly twisted instead of lying
completely in-plane with the skeleton of DO11 as shown in Figure \ref{fig:DFTsturctures}.

The IR absorption peaks of oxetane in the gas phase are located at
900 cm\textsuperscript{-1}, 995 cm\textsuperscript{-1}, and 2850$\sim$3050
cm\textsuperscript{-1} \cite{NISToxetane}. The characteristic IR
band of oxetane formed between 1AAQ and olefines was observed at 990
cm\textsuperscript{-1}, and for oxetane formed between 1AAQ and styrene
was found at 970 cm\textsuperscript{-1} \cite{Inoue1982b}. In this
study, though, the change in the IR spectrum is relatively noisy in
the frequency range below 1000 cm\textsuperscript{-1} and no significant
change is observed between 800 cm\textsuperscript{-1} and 1000 cm\textsuperscript{-1}
after irradiation. Since the vibration of chemical bonds may be affected
by the environment, the growing peak at 1136 cm\textsuperscript{-1}
after irradiation may be attributed to oxetane formed between 1AAQ
and MMA as shown in Figure \ref{fig:140729ex2Y105MftirPTCR}(d). The
recovery of the peak at 1136 cm\textsuperscript{-1} indicates that
1AAQ-MMA oxetane molecules return to 1AAQ and MMA or further decay
into irreversibly-damaged species. Alternatively, the peak at 1136
cm\textsuperscript{-1} may result from a frequency shift of the 1150
cm\textsuperscript{-1} peak due to the dissociation of the \ce{COOCH3}
group from PMMA under irradiation and recovers after the laser is
turned off. In this case, another photochemical reaction other than
photocycloaddition between 1AAQ and MMA may occur resulting in a reversibly-damaged
dye species since the vibration frequency of oxetane is not observed. 
Nonetheless, the reversible damage of 1AAQ occurring at
the carbonyl group adjacent to the amine group is observed.

The change in the IR absorption peak at 1280 cm\textsuperscript{-1}, which corresponds
to the \ce{C-C-C} stretching vibration of ketone adjacent to
the amine group in 1AAQ, partially recovers and is distinguishable 
until the 540\textsuperscript{th}
minute. However, a growing peak at 1280 cm\textsuperscript{-1}
is also observed at long-times as described
in Section B of supplemental material. Thus, the recovery of the peak at 1280 cm\textsuperscript{-1}
may have become indistinguishable from noise at the
150\textsuperscript{th} minute. It is difficult to determine the
time that the recovery of the in-plane \ce{NH2} scissoring band
at 1546 cm\textsuperscript{-1} and the carbonyl group stretching
band at 1608 cm\textsuperscript{-1} becomes indistinguishable from
noise due to the small signal-to-noise (S/N) ratio of the spectrum.
While partial recovery of 1AAQ agrees with the observation in
absorption spectroscopy measurements, the reduced IR peaks belonging
to PMMA fully recover at the 150\textsuperscript{th} minute. The
disagreement of recovery dynamics between 1AAQ and PMMA may originate from
the changes in the background of the system due to contamination as described in Section
B of supplemental material, the small S/N ratio of the spectrum, and uncertainties and artifacts
that arise from the baseline process used for obtaining the spectra.
Nonetheless, the experimental results qualitatively agree with the
PTCR hypothesis.

The PTCR hypothesis is consistent with the spectra predicted using TD-DFT calculations,
the measured absorption spectra of the pristine and irreversibly-damaged species, the absorption spectrum of the reversibly-damaged species obtained using a simple three species model, and changes in molecular structures of dye and polymer upon photodegradation and during recovery observed in the FTIR experiment in this study. The chemical
reactions involved in the PTCR hypothesis suggest that the decay and
recovery processes may be governed by energy barriers among pristine
dye and degraded dye species. However, the energy barrier scenario
seems to contradict a phenomenological model that is based on quantitative
studies of reversible photodegradation in DO11/PMMA using various
experimental techniques \cite{Ramini2012b,Ramini2013,Anderson2014a}.
The kinetics of the decay and recovery
processes are currently under investigation to better understand the possible causes of the disagreement as briefly discussed elsewhere \cite{Hung2015}.

\section{Conclusion}
The mechanism responsible for reversible photodegradation in 1-substitued
aminoanthraquinone-doped polymers has been studied using GPC, TD-DFT calculatioins,
UV-Vis and FTIR spectroscopy. GPC experiments eliminate the possibility of 
residual monomers being responsible for the observed reversible photodegradation  
in dye-doped polymers. TD-DFT calculations indicate that the
reversible and irreversible degradation in the dye molecule occur
at the carbonyl group adjacent to the amine group, and qualitatively
agrees with the experimental results obtained using UV-Vis and FTIR
spectroscopy. Studies using FTIR spectroscopy also indicates that both
dye and polymer undergo reversible photodegradation, instead of the dye
alone. These results suggest that photodegradation is caused
by (photo)chemical reactions between pristine dye and the photoinduced thermally-degraded
polymer, and recovery is due to the metastable (photo)chemical reaction
product returning back to pristine dye.

The observed photoinduced thermal degradation may occur in various
doped polymers, and thus (photo)chemical reactions between dopants
and thermally-degraded polymers may be one common mechanism responsible
for reversible and irreversible photodegradation depending on the
stability of the reaction products. Thus, photoinduced thermal degradation
in polymers is a factor that should be taken into account,
in addition to the effects of oxygen, moisture etc., to understand how to improve the
stability of doped polymer materials when light exposure is required. Methods such as reducing the
probability of non-radiative relaxation from dopants, increasing the
thermal degradation temperature of the polymer host, increasing the
heat dissipation rate of polymer chains, selecting proper dopants
that do not react with thermally degraded polymers under the operating
condition (if local thermal degradation of polymer chains is not a
concern) etc. are possible approaches to improve photostability.

Although energy transfer between dye and polymer is the key to photodegradation, the mechanism of the energy transfer is not clear. Understanding the mechanism of energy transfer between dye and polymer will provide a new perspective to design desirable dye-doped polymer materials. For example, F{\"o}rster resonance energy transfer theory has been applied to simulate the dynamics of dye excitation and energy transfer in nanostructured DNA-dye assemblies for light-harvesting applications \cite{Pan2014}. The study in kinetics of reversible photodegradation of dye-doped polymers is currently ongoing. The available results so far suggest that the local environment of dye molecules drastically affects the photodegradation and recovery rate \cite{Hung2015}, hence the energy transfer between dye and polymer and the involved (photo)chemical reactions may greatly depend on the distribution of dye in the polymer host. The knowledge of energy transfer mechanism and dye distribution in the polymer can be applied to design dye-doped polymers with improved device performance and photostability.

\section{Acknowledgments}
S.T.H. and M.G.K. thank Prof. Kirk A. Peterson for valuable help and
advice in TD-DFT calculations, and Wright-Patterson Air Force Base
and Air Force Office of Scientific Research (FA9550-
10- 1-0286) for their support of this
research. S.T.H. thanks Dr. Candy C. Mercado for valuable discussion
in molecular spectroscopy, and WSU for a College of Arts and Sciences
Research Assistantship.


\bibliographystyle{osajnl}
\bibliography{allpapers}

\end{document}